\documentclass[10pt]{article}

\pdfoutput=1

\usepackage[textwidth=15.5cm,textheight=22.5cm]{geometry}
\usepackage{bm}
\usepackage{amsmath,amssymb,color,graphicx,amscd,amsfonts}
\usepackage{latexsym}
\usepackage{graphicx}
\usepackage{cite}
\usepackage{subfig}
\usepackage{bbm}
\usepackage{ulem}
\usepackage[usenames,dvipsnames,svgnames,table]{xcolor}

\usepackage{datetime}

\newcommand{\rc}{r_{\textrm{bunch}}}
\newcommand{\rt}{r_{\textrm{trap}}}
\newcommand{\rp}{r_{\textrm{peak}}}
\newcommand{\rh}{r_{\textrm{horizon}}}

\begin{document}

\thispagestyle{empty}

\setcounter{page}{0}

\begin{flushright} 

SU-ITP-17-05\\
YITP-17-55\\
LLNL-JRNL-732009\\
RBRC-1251
\end{flushright} 

\vspace{0.1cm}

\begin{center}
{\LARGE

Toward Holographic Reconstruction of Bulk Geometry\\
from Lattice Simulations

\rule{0pt}{20pt}  }
\end{center}

\vspace*{0.2cm}

\renewcommand{\thefootnote}{\alph{footnote}}

\begin{center}
	Enrico R{\sc inaldi}$^{abh}$,	
	Evan B{\sc erkowitz}$^{ac}$,
	Masanori H{\sc anada}$^{adef}$, 
          Jonathan M{\sc altz}$^{dg}$,\\
           and Pavlos V{\sc ranas}$^{ah}$

\vspace{0.3cm}
$^a$ {\it Nuclear and Chemical Sciences Division, Lawrence Livermore National Laboratory, \\
Livermore CA 94550, USA}

$^b$
{\it RIKEN-BNL Research Center, Brookhaven National Laboratory, Upton, NY 11973, USA}

$^c$ {\it Institut f\"{u}r Kernphysik and Institute for Advanced Simulation, \\
Forschungszentrum J\"{u}lich, 52425 J\"{u}lich, Germany}

$^d${\it Stanford Institute for Theoretical Physics, Stanford University, Stanford, CA 94305, USA}

$^e${\it Yukawa Institute for Theoretical Physics, Kyoto University,\\
Kitashirakawa Oiwakecho, Sakyo-ku, Kyoto 606-8502, Japan},

$^f$	
{\it The Hakubi Center for Advanced Research, Kyoto University,\\
Yoshida Ushinomiyacho, Sakyo-ku, Kyoto 606-8501, Japan}

 $^g$
{\it Berkeley Center for Theoretical Physics, University of California at Berkeley,\\
 Berkeley, CA 94720, USA}

$^h$
{\it Nuclear Science Division, Lawrence Berkeley National Laboratory, \\
Berkeley, CA 94720, USA}

erinaldi@bnl.gov, e.berkowitz@fz-juelich.de, hanada@yukawa.kyoto-u.ac.jp, jdmaltz@berkeley.edu, vranas2@llnl.gov

\end{center}

\vspace{1cm}

\begin{abstract}

A black hole described in SU($N$) gauge theory consists of $N$ D-branes. 
By separating one of the D-branes from others and studying the interaction between them,
the black hole geometry can be probed.
In order to obtain quantitative results, we employ the lattice Monte Carlo simulation.
As a proof of the concept, 
we perform an explicit calculation in the matrix model dual to the black zero-brane in type IIA string theory. 
We demonstrate this method actually works in the high temperature region, where the stringy correction is large. We argue possible dual gravity interpretations.

\end{abstract}

\newpage
\section{Introduction}
\hspace{0.25in}

How bulk spacetime emerges from the dual gauge theory has been one of the most important areas of study in quantum gravity.
Historically, large-$N$ volume reduction \cite{Eguchi:1982nm} demonstrated that spacetime can be encoded in matrix degrees of freedom. 
In this context, it has been realized that the eigenvalue distribution of the matrices is closely related to the geometry \cite{Bhanot:1982sh,Gross:1982at,Parisi:1982gp,GonzalezArroyo:1982hz}. 
From the point of view of superstring/M-theory, eigenvalues correspond to the positions of D-branes and various objects can be constructed simply as bound states of D-branes and open strings\cite{Witten:1995im,Banks:1996vh}. 
The large-$N$ volume reduction is then analogous to the emergence of higher dimensional branes from lower dimensional branes, e.g. D($p+2$)-branes from D($p$)-branes via the Myers effect\cite{Myers:1999ps}. 

Although such approaches have been successful for various purposes, they have not sufficiently demonstrated how to understand the emergent geometry in holography \cite{'tHooft:1993gx,Susskind:1994vu,Maldacena:1997re} because it is necessary to understand how dimensions transverse to the branes emerge. 
Most of the recent studies concentrate on conformal field theories dual to AdS spaces, and consider the construction of bulk local operators from non-local operators on the boundary \cite{Banks:1998dd,Hamilton:2006az,Heemskerk:2009pn,Heemskerk:2012mn}. 
In this paper, we propose---or rather, revisit---a simple method, which is (at least seemingly) different 
and applicable to more generic theories.
In fact, our strategy is very straightforward: we follow the old interpretation of Refs. \cite{Witten:1995im,Banks:1996vh,deWit:1988ig}, and we solve the dynamics of gauge theory from first principles.
 
In the Matrix Theory proposal \cite{Banks:1996vh}, gravitational interactions are obtained from the interactions between D0-branes.  
Therefore, by looking at the interactions in a specific system of D0-branes --- forming an extended object such as a black hole --- together with a ``probe'' D0-brane whose position is moved by hand, it is possible to obtain the information about the geometry as the force acting on this probe.
The same idea applies to any gauge theory which has D-brane origins, and has also played an important role for the discovery of gauge/gravity duality (see e.g. \cite{Maldacena:1997nx,Tseytlin:1998cq}). 
In particular, the eigenvalues are expected to be described by the Dirac-Born-Infeld action \cite{Maldacena:1997re}, from which the spacetime geometry can be reconstructed. 
Further studies along this line and related directions include Refs. \cite{Dorey:1999pd,Buchbinder:2001ui,Iizuka:2001cw,Kuzenko:2004sv,Ferrari:2012nw,Buchbinder:2016xeq,Schwarz:2014zsa,Sahakian:2017uwm}. 
Moreover, a similar idea has been studied in the context of entanglement entropy \cite{Mollabashi:2014qfa}, in order to see how the $\mathbb{S}^5$ of AdS$^5\times\mathbb{S}^5$ geometry emerges.

In the past, there have also been attempts \cite{Berenstein:2013tya} to study the ``internal'' structure in a system of D0-branes, focusing on a region of parameter space (at high temperatures) where classical or \emph{semi}-classical approaches are a good approximation for the dynamics.
Our focus is a regime of temperatures where the gauge theory is strongly coupled, and, even though the probe brane approach described in this work is intuitively simple, it remains challenging because of the obvious difficulties in calculating observables non-perturbatively.
In this paper, we employ numerical Monte Carlo methods to overcome this difficulty in the strongly coupled regime.

This paper is organized as follows. 
In Sec.~\ref{sec:proposal} we consider the dynamics on the gauge theory side. 
Although the quantitative calculation is hard, without relying on numerics, a qualitative picture of the gauge theory calculation is provided. 
Physics captured by classical studies \cite{Berenstein:2013tya} and features added by the new full quantum treatment will be explained. 
In Sec.~\ref{sec:numerics} we use numerical Monte Carlo method to confirm this picture. 
In Sec.~\ref{sec:gravity_dual}, we list possible dual gravity interpretations of the calculation. 
Note that the parameter region we numerically studied corresponds to a rather stringy regime on the gravity side, and hence the dual gravity interpretation can be speculative.

\section{The Gauge Theory Picture}\label{sec:proposal}
\hspace{0.25in}

In this section, we describe the proposed method to investigate how the black hole geometry can be detected directly in the gauge theory. 
Before discussing various interpretations of the dual gravity theory, we define the problem at hand in the gauge theory picture. 

As a concrete example, let us consider the matrix model of D0-branes\footnote{Generalizations to higher dimensions are straightforward.}, 
which is numerically tractable with reasonable computational resources.
The Lagrangian of the theory is 
\begin{eqnarray}
\mathcal{L}
&=&
\frac{1}{2g_{YM}^2}{\rm Tr}\Bigg\{
(D_t X_M)^2 
+
[X_M,X_{M'}]^2 
+
i\bar{\psi}^\alpha D_t\psi^\beta
+
\bar{\psi}^\alpha\gamma^M_{\alpha\beta}[X_M,\psi^\beta] 
\Bigg\},  
\end{eqnarray}
where $X_M$ $(M=1,2,\cdots,9)$ are $N\times N$ Hermitian matrices and $(D_tX_M)$ is the covariant derivative given by $(D_tX_M)=\partial_t X_M-i[A_t,X_M]$ and $A_t$ is the $U(N)$ gauge field.
The gamma matrices $\gamma^M_{\alpha\beta}$ $(M=1,2,\cdots,9)$ are the $16\times 16$ left-handed part of the gamma matrices in ($9+1$)-dimensions. 
$\psi_\alpha$ $(\alpha=1,2,\cdots,16)$ are $N\times N$ real fermionic matrices.
This Lagrangian is the dimensional reduction of 4D ${\cal N}=4$ super Yang-Mills theory to ($0+1$)-dimensions. 

We set the 't Hooft coupling $\lambda=g_{YM}^2N$ to one unless $\lambda$ is explicitly shown. Equivalently, all dimensionful quantities are measured in units of the 't Hooft coupling; 
for example the temperature $T$ actually refers to the dimensionless combination $\lambda^{-1/3}T$.

In this section we will consider the micro-canonical ensemble in the theory with Minkowski signature, since we will eventually be interested in the black hole geometry in Minkowski space.
When interpreted as the low-energy effective description of open strings and D0-branes, the diagonal and off-diagonal elements of $X_M$ can be regarded as the D0-branes and open strings, respectively \cite{Witten:1995im}. 
This theory can describe multiple objects (such as multi-graviton or black hole states) through block-diagonal matrices, where each block corresponds to a different object \cite{Banks:1996vh}.
Interactions are then mediated by the quantum fluctuations of off-diagonal elements.

Let us consider a typical matrix configuration about the trivial vacuum, which is a bunch of $N$ D0-branes.
Separating one of the D0-branes, which is represented by the ($N,N$)-element of $X_M$, from the bunch allows us to regard 
this D0-brane as a probe\footnote{
More precisely, we take the $A_t=0$ gauge, in which the structure of the physical Hilbert space has a natural connection to open strings and D0-branes, and then separate the ($N,N$)-component from the others. 
}.
The matrices are then of the form,  
\begin{eqnarray}
X^M
=
\left(
\begin{array}{cc}
X_{\rm BH}^M & w^M \\
w^{\dagger M} & x_{\rm D0}^M
\end{array}
\right), \label{eq:decomposition}
\end{eqnarray}
where $w^M$ describes a small fluctuation of the $N$-th row and column, which are interpreted as open string excitations between the probe $x_{\rm D0}^M$ and the rest of the original bunch $X_{\rm BH}^M$.\footnote{Here we have used the subscript ``BH'' because, later in this paper, we will interpret the bunch as a black hole (black zero-brane) via gauge/gravity duality.}

When we interpret the diagonal element $x_{\rm D0}^M$ to be the position of a D0-brane, we implicitly assume that the off-diagonal elements $w^M$ are small.
One possible criterion for the smallness of $w^M$, which we will adopt in this paper, is that $O(w^3)$ terms of the action are negligible and $w^M$ behaves as a harmonic oscillator. When $w$ is so large that $O(w^3)$ terms are no longer negligible, corrections to this simple geometric picture \cite{Witten:1995im} will be needed.
Note that, even when no open string is excited, $|w|$ cannot be exactly zero, due to the zero-point oscillations of the harmonic oscillator. 
The zero-point fluctuations become large when the probe gets close to the bunch. 
Hence, even at zero temperature, the off-diagonal elements become large at short distances and it is probably not appropriate to interpret the diagonal elements as the positions of D0-branes. 
The crossover between these two regimes takes place when $T\sim 1$.

One subtle point associated with such zero-point fluctuations is the interpretation of the bunch, $X_{\rm BH}$. 
It is highly non-commutative at any temperature. 
At high temperatures, the non-commutativity is dominated by thermal excitations of open strings, which invalidate a classical geometric picture on the gravity side.
On the other hand, at sufficiently low temperatures, the main source of the non-commutativity are the ``zero-point oscillations''; 
then, while the diagonal elements may not be the positions of D0-branes\footnote{
At short distances, higher order terms can contribute and the off-diagonal elements do not behave as decoupled harmonic oscillators. Due to this, the  simple ``zero-point fluctuations'' picture may not be appropriate. 
However, in \cite{Azeyanagi:2009zf}, in a similar theory (possessing 4 supercharges rather than 16), it was numerically observed that the higher order terms 
give only small contributions and ``zero-point fluctuations'' picture is rather good. 
} , 
the classical geometry on the gravity side may still make sense because there are no open string excitations. 
In this paper we study only $T\gtrsim 1$, because the crossover between these two regimes takes place at $T\sim 1$.  

Our approach in this paper is to define the \emph{distance} between the probe D0-brane $x_{\rm D0}^M$ and the center of the bunch $\frac{{\rm Tr}X_{\rm BH}^M}{N-1}$ as
\begin{equation}
  \label{eq:distance}
r \equiv \sqrt{\sum_M\left( \frac{{\rm Tr}X_{\rm BH}^M}{N-1}-x_{\rm D0}^M \right)^2} \quad ,
\end{equation}
and then numerically calculate the force applied by the bunch on the probe as a function of this particular distance.
This allows us to obtain insights of the geometry from the dual gravity picture. 
Note that we take the large-$N$ limit for a fixed value of $r$ and therefore $w^M$ and $x_{\rm D0}^M$ can be treated as a ``subsystem'' interacting with a thermal bath described by $X_{\rm BH}^M$. 

\begin{figure}[htbp]
\begin{center}
\rotatebox{0}{
\includegraphics[scale=0.5]{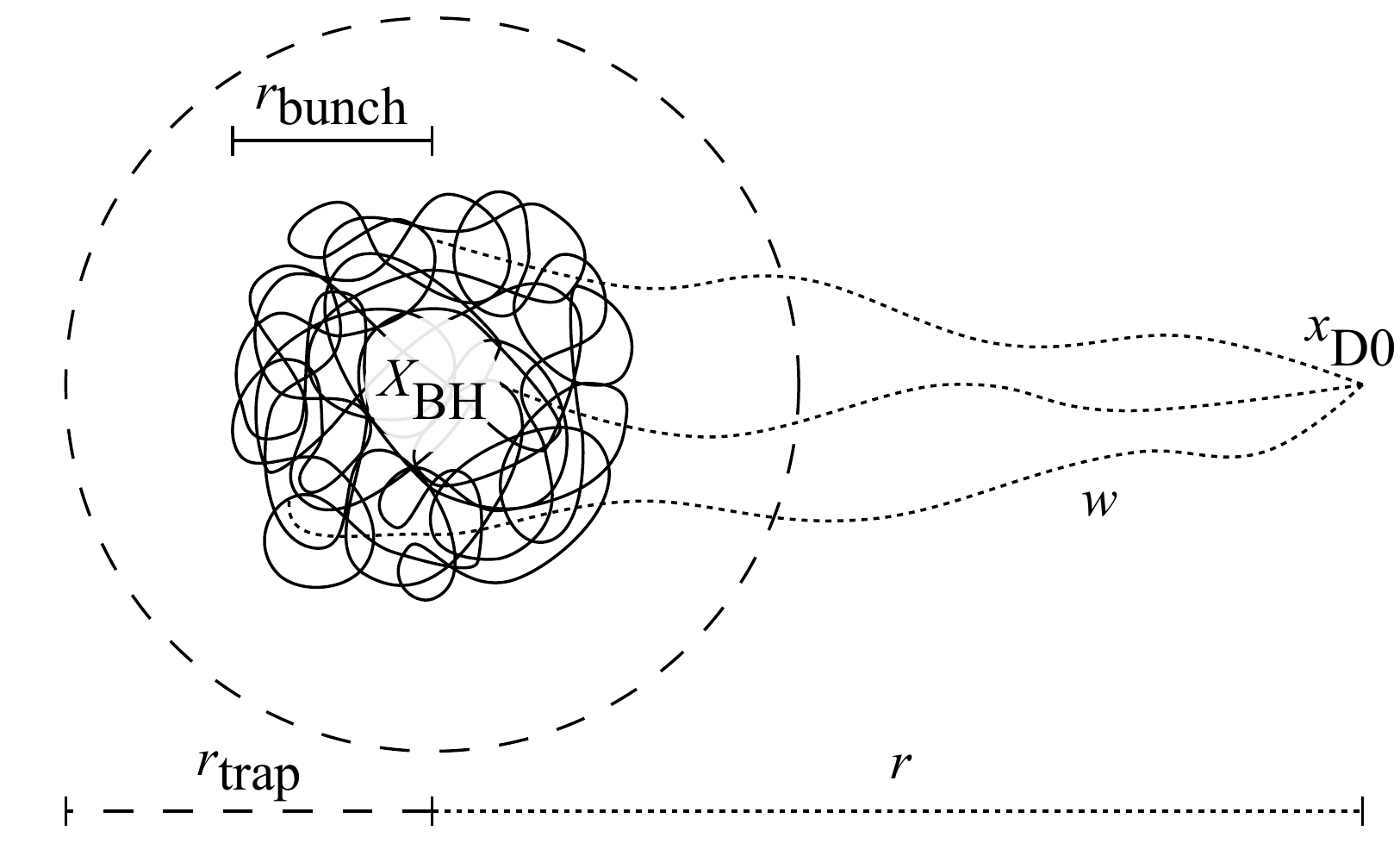}}
\end{center}
\caption{A ``black hole geometry'' in a gauge theory. We also show the probe at $x_{\rm D0}$ and the open strings $w$ that connect it to the black hole $X_{BH}$, and the length scales $r$, $\rt$, and $\rc$.
}\label{fig:horizon}
\end{figure}

As the distance between the bunch and the probe varies, the force should behave as follows (see Fig.~\ref{fig:horizon}\footnote{This is a refinement of the idea suggested in \cite{Berkowitz:2016znt,essay}.}):

\begin{itemize}

\item

{\bf Short distance:} The probe merges into the bunch of other D-branes.
The off-diagonal elements $w_M$ and $w_M^\dagger$ condense and the ``position'' of the probe can not be defined in a meaningful sense\footnote{This is similar to the phase transition in a related model studied in \cite{Ferrari:2016bvq}.}.
We call this region the ``bunch''.
The radius of the bunch can be estimated as
$\rc \equiv
\sqrt{
\left\langle
\frac{1}{N}
\sum_{M=1}^9{\rm Tr}(X^M)^2
\right\rangle}$ when the probe is absent, or using
$\rc\equiv\sqrt{\left\langle\frac{1}{N-1}\sum_{M=1}^9{\rm Tr}(X_{\rm BH}^M)^2\right\rangle}$.
The latter is a good estimate when $N$ is large and it can be used in the presence of the probe, with the caveat that the distribution in the 9-dimensional space will be skewed in the direction of the probe when the acting force is large (we will comment on this later).
In this paper we will use these two definitions interchangeably. 

As we have mentioned above, $\rc$ is non-zero even at $T=0$, due to quantum fluctuations. 
At sufficiently low temperatures the classical geometry on the gravity side may make sense even inside the bunch. 

\item

{\bf Long distance:} The force goes as $\sim f(T)\cdot Nr^{-8}$\cite{Banks:1996vh,Danielsson:1996uw,Kabat:1996cu}, where the temperature-dependent prefactor $f(T)$ disappears at $T=0$.

\item
{\bf Intermediate distance:} Here is where non-trivial dynamics can emerge. 
Firstly, off-diagonal elements are not very large and the position of the probe makes approximate sense.
As the probe approaches the bunch, open string excitations become increasingly important and numerical calculations of the force are required in order to understand this region. 
This is also where perturbative analysis is expected not to work.
We expect the shape of the bunch to deform in response to the probe.
In analogy with the Moon's tidal effect on the Earth's oceans, we expect the bunch to become prolate.

\end{itemize}

In order to obtain a better picture of the dynamics at intermediate distances, let us consider the
$T\gg 1$ regime where, on the gravity side, $\alpha'$ corrections will become important.

\begin{itemize}
\item
When $r \lesssim T$, off-diagonal elements are highly excited and non-perturbative effects become important.
A strong attractive force is expected.\footnote{
The same dynamics has been discussed in \cite{Kofman:2004yc} 
as `moduli trapping'. 
} 

\item
When $r \gtrsim T$, the off-diagonal elements are exponentially suppressed as they are too heavy and decoupled from the dynamics, making the one-loop approximation valid.

\item
The size of the bunch scales as $\rc \sim T^{1/4}$, see e.g. Refs.\cite{Berkowitz:2016znt,essay}. 
Therefore, the intermediate distance region is separated into two parts: $T^{1/4}\lesssim r \lesssim T$ and $r\gtrsim T$. 
The emission of eigenvalues from $r \lesssim T$ is entropically suppressed with a suppression factor $\sim e^{-N}$, because an $O(N)$ number of off-diagonal elements must be suppressed simultaneously \cite{Berkowitz:2016znt,essay}\footnote{ 
The emission's suppression is also understood as follows. 
As we will demonstrate numerically in Sec.~\ref{sec:numerics}, the attractive force is of order $N$. 
The mass of the brane is of order $N$, and hence the D0-brane must have an order one velocity in order to escape. 
However, the typical energy and velocity are of order 1 and $1/\sqrt{N}$, respectively, 
because the energy is of order $N^2$ and there are order $N^2$ degrees of freedom including the open strings.  
}. 
In the large-$N$ limit, the eigenvalues cannot escape once they reach the region $r\lesssim T$.
Following this reasoning, we call $r \sim T$ the {\it trapping radius} and denote this distance by $\rt$. 
A schematic representation of the various distances at play is shown in Fig.~\ref{fig:horizon}. 

\end{itemize}


Classical simulations (e.g. see Ref.~\cite{Berenstein:2013tya}) should be a valid approximation to the full quantum theory at $r\ll T$.
Therefore, physics near the bunch, for example the thermalization of a black hole \cite{Asplund:2011qj}, can be understood based on results from classical simulations. 
However, the classical approximation breaks down at $r \gtrsim T$, because the mass of strings -- the energy quanta -- becomes non-negligible compared to the energy scale $T$. 
In this region, for example, quantum effects assist the evaporation of a black hole \cite{Berkowitz:2016znt,essay}.  

The high-temperature picture should fail for $T \lesssim 1$, where $\rc$ and $\rt$ become of the same order. 
Below that point, we can immediately imagine two natural possibilities: either $\rt$ approaches $\rc$ and they coincide at $T=0$ or $\rt$ coincides with $\rc$ at finite temperature. 
Regardless of the relationship between $\rt$ and $\rc$, the force acting on the probe should cancel at $T=0$ due to supersymmetry. 

In the rest of the paper we will show that our numerical results are consistent with these expectations.

\section{The Numerical Demonstration}\label{sec:numerics}
\hspace{0.25in}

In this section, we demonstrate the scenario described above by performing explicit calculations in the gauge theory. 
Although the force depends on the relative velocity between the black hole and the probe, we will concentrate on the case with zero relative velocity for a practical reason explained below. 

To begin, we modify the potential by adding terms which will fix the distance between $X_{\rm BH}^M$ and $x_{\rm D0}^M$, up to quantum fluctuations.
If the black hole is not spinning then by rotational symmetry we can take the displacement of the probe to be along the $M=1$ direction. 
We add\footnote{
This deformation manifestly breaks U$(N)$ to U$(N-1)\times$U$(1)$. 
In principle, we can make a gauge-invariant analogue of this deformed potential, for example by fixing the position of the largest eigenvalue of $X_1$. We chose this specific deformation because it is technically easy. 
} to the action
\begin{equation}
\Delta \mathcal{L} =
-c\Bigl\{
\left(\frac{{\rm Tr}X_{\rm BH}^1}{N-1}-x^1_{\rm D0}-r_0\right)^2
+
\sum_{M=2}^9
\left(\frac{{\rm Tr}X_{\rm BH}^M}{N-1}-x^M_{\rm D0}\right)^2
\Bigl\}
-c'|w_1|^2
\label{eq:deformation}
\end{equation}
to the Lagrangian, in order to hold the probe D0-brane
near the position $R=(r_0,\vec{0})$, where $r_0$ is the coordinate in the $M=1$ direction and $\vec{0}$ is an eight-dimensional vector.
Hence, we are introducing three new parameters, $\{ c,c',r_0 \}$. The last one, $r_0$, is fixed in each simulation to constrain the distance of the probe, while we vary the first two in order to check that we are in a regime where the final results are unaffected by our choice (within our total statistical uncertainty).
%
In particular, the last term is needed in order to remove the unphysical longitudinal oscillation modes of the open strings, and the value of $c'$ is taken to be rather large $\sim 100$, and fixed throughout our simulations.

An important remark is that, because of the interaction mediated by the off-diagonal elements and the quantum mechanical nature of the system, the measured distance according to our definition in \eqref{eq:distance} will deviate from $r_0$ in the $M=1$ direction (and also slightly in the other directions).
In the numerical simulation we measure the following expectation values
\begin{align}
  \label{eq:observables}
  r_{M=1} & \equiv \left\langle \frac{{\rm Tr}X_{\rm BH}^1}{N-1}-x^1_{\rm D0} \right\rangle \nonumber  \\
  r_{M=2} &  \equiv \left\langle \frac{{\rm Tr}X_{\rm BH}^2}{N-1}-x^2_{\rm D0} \right\rangle \quad ,
\end{align}
where the second distance, which should be distributed around zero, is only used as a cross-check to monitor that the deformation in \eqref{eq:deformation} is working as expected.
In all our simulations, with varying values of $c$, we find $r_{M=2} \approx 0$ and therefore we identify the distance in \eqref{eq:distance} with $r_{M=1}$.
At distance $r$, the force $F$ between the probe and the bunch is canceled by the additional force coming from $\Delta \mathcal{L}$.
Therefore we can define a force for each value of the input parameters, $N$, $r_0$ and $c$, as
\begin{equation}
  \label{eq:force}
  F(N,r_0;c) \; = \; 2c(r_0-r) \ ,
\end{equation}
 up to higher order terms in $r_0-r$, where, again, $r$ is our primary observable that we identify with $r_{M=1}$ in \eqref{eq:observables}.
Although this should be interpreted as the force at distance $r$, we took $c$ sufficiently large so that $r$ and $r_0$ are always very close.
Hence we will regard it as the force at distance $r_0$ when we show it later in the paper.
In Appendix~\ref{sec:force-calculation} we show a typical example of our numerical simulations and we show the measured observables to demonstrate in details all the points above.

Note that the force calculated in this manner does not contain the effect of the velocity of the probe.
Note also that the deformation on the dual gravity theory caused by this additional deformation term is not clear.
We have introduced $\Delta \mathcal{L}$ only as a trick to determine the force on the gauge theory side. 
When we discuss the dual gravity interpretation, we will only consider the standard duality, in the absence of this modification term.

With this deformation $\Delta \mathcal{L}$, the configuration is made static.
Therefore we can Wick rotate the system to Euclidean signature in order to measure the force\footnote{
In the Euclidean theory at finite temperature, the gauge field $A_t$ cannot be set to zero.
Instead we have used the static diagonal gauge, $A_t={\rm diag}(\alpha_1,\alpha_2,\cdots,\alpha_N)$, where $\alpha_i$'s are $t$-independent and satisfy $0\le \alpha_i < 2\pi T$. 
}.
We perform the path integral in imaginary time by using Monte Carlo methods,\footnote{Ofer Aharony suggested this numerical experiment to M.~H. in 2009. 
At that time M.~H. did not try it because the physical picture was not clear to him. 
M.~H. thanks Ofer Aharony for the valuable advice.}
so the result obtained corresponds to the canonical ensemble. 
At large-$N$, this should give the same result as the micro-canonical ensemble. 

We added the deformation term \eqref{eq:deformation} to a lattice simulation code for the Monte Carlo String/M-theory Collaboration \cite{simulation_code}.  
We studied $T=1.0, 1.5, 2.0$ and $3.0$ for matrices in SU($N$) with $N=6$ to $N=16$, and with a variety of lattice spacings determined by $L$, 
going from $L=8$ to $L=24$.
The 't Hooft coupling $\lambda=g_{YM}^2N$ is set to $1$.

In Fig.~\ref{fig:force_plot} we show the normalized force $F/(N-1)$ as a function of the position $r_0$.
 The two panels correspond to two different temperatures, $T=1.0$ and $T=2.0$, and numerical results with $N=6,8,12$ and $16$ are included.
Note that for the largest value of $N$, we do not have results around the peak of the force.
From this numerical data of the force, we can identify interesting features pertaining to different distance regimes.

\begin{figure}[htbp]
\begin{center}
\rotatebox{0}{
\scalebox{0.48}{
\includegraphics{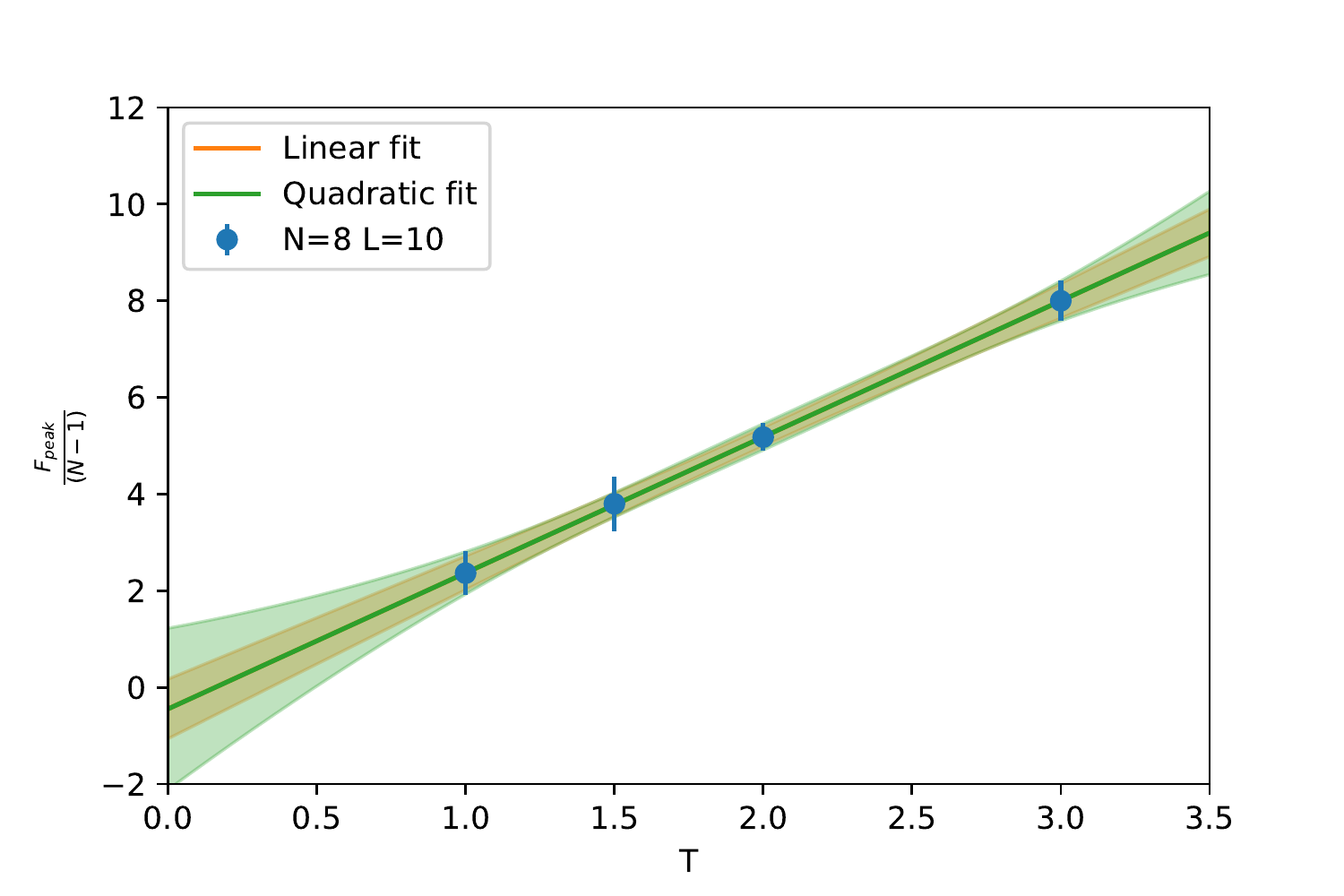}}}
\rotatebox{0}{
\scalebox{0.48}{
\includegraphics{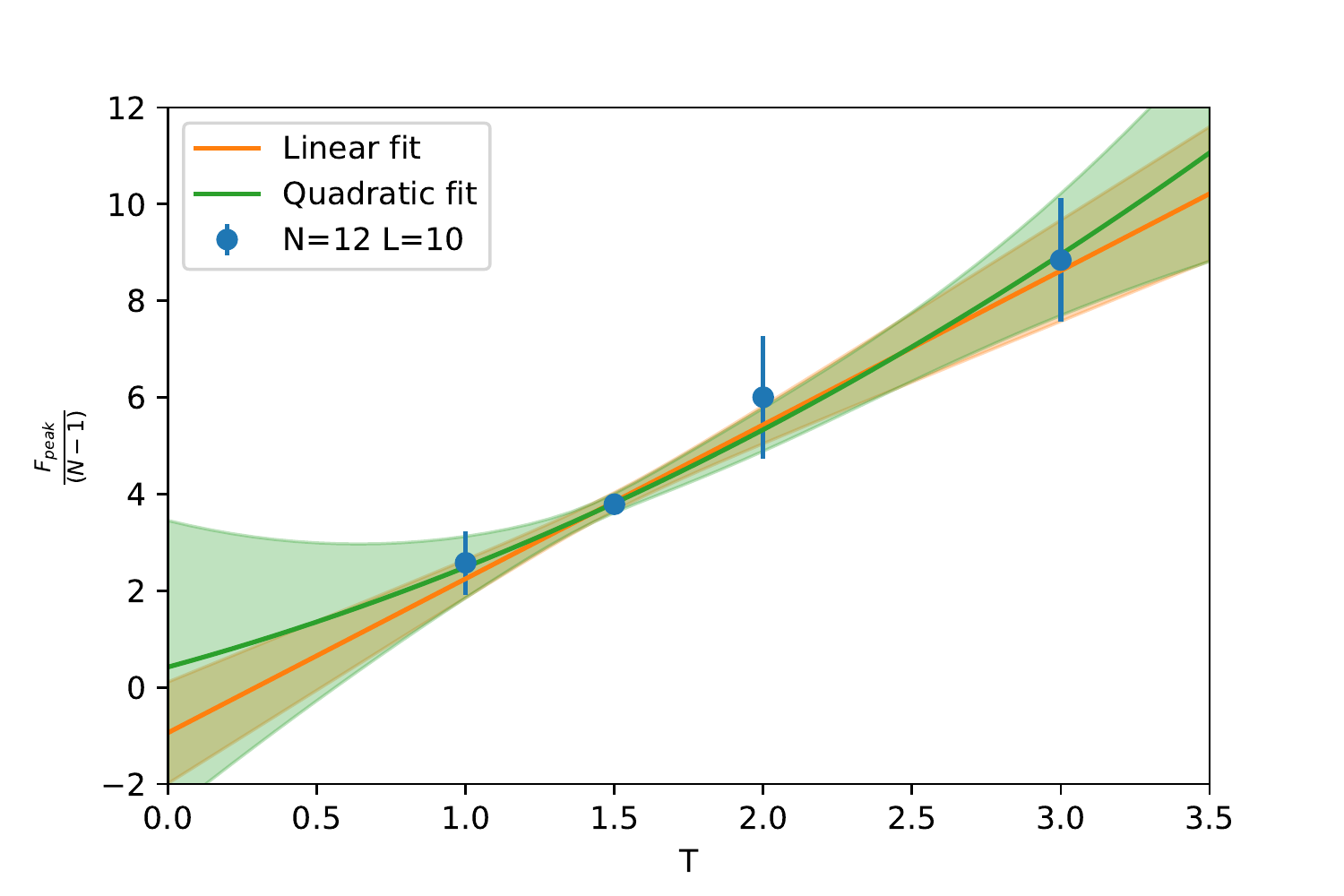}}}
\end{center}
\caption{
The largest value of the force, $F_{\rm peak}/(N-1)$, with $N=8$ and $N=12$ at fixed lattice spacing $L=10$. 
}\label{fig:force-height}
\end{figure}

At short distance, $F/(N-1)$ takes positive values, which confirms the ${\cal O}(N)$ attraction region described in the previous section.
There is a peak at some distance $\rp$, which we numerically determined as the interval encompassing the three largest values of the force.
We also estimate the value of the maximal force $F_{\rm peak}/(N-1)$ and its systematic uncertainty, due to finite $r_0$ spacing, using the distance between the maximum and the third largest force (note that the statistical error is always much smaller than this systematic error).
The maximal force is shown in Fig.~\ref{fig:force-height} as a function of the temperature for $N=8$ and $N=10$ at fixed lattice spacing $L=10$.
Simple extrapolations using linear and quadratic ans\"{a}tze indicate that the data is consistent with a null maximal force at $T=0$.

\begin{figure}[htbp]
\begin{center}
\rotatebox{0}{
\scalebox{0.3}{
\includegraphics{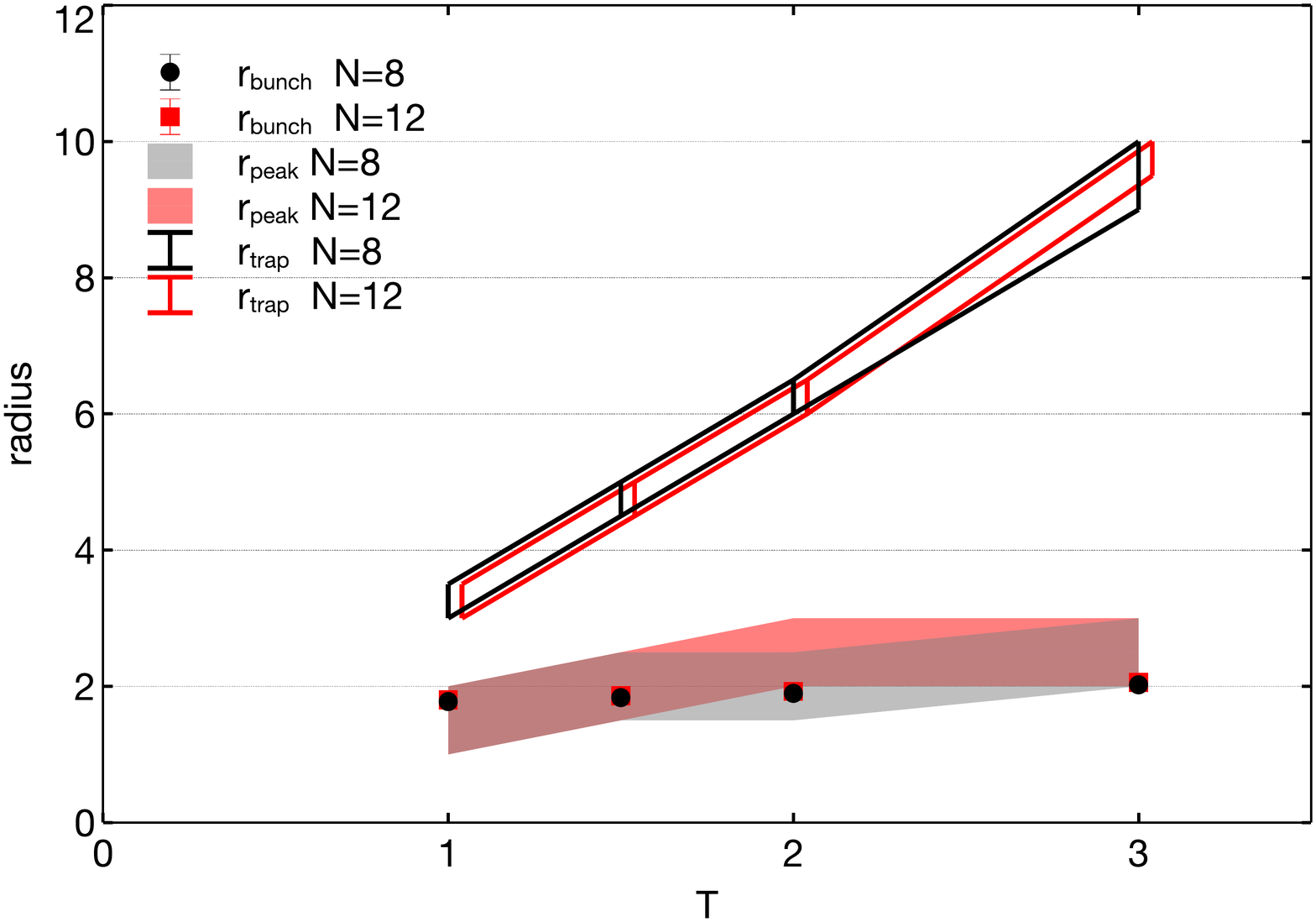}}}
\end{center}
\caption{
Plot of $\rc$, $\rp$ and $\rt$ for $L=10$ and two values of $N$. The $\rt$ results for different $N$ values are horizontally displaced for clarity.
The values of $\rp$ and $\rt$ are only determined as intervals between different simulated values of $r_0$: the former is determined by the interval containing the three largest values of the force, while the latter is determined by subsequent values of $r_0$ where the force changes sign from positive to negative.}
\label{fig:rcrh}
\end{figure}

In Fig.~\ref{fig:rcrh} we summarize the various distances, or ``radii'', at play in the system, for $N=8,12$ and $L=10$.
 We can see, for example, that the peak of the force $\rp$ coincides, within uncertainties, with $\rc$.
This suggests that the force decreases once the probe merges into the bunch.
It is easy to understand this feature of the data: when the probe approaches the origin from the right on the positive $x^1$ side, outside the bunch, the probe is pulled only to the left, while in the bunch some D0-branes pull the probe in the opposite direction.
At the center of the bunch $r_0\approx 0$, the force should cancel due to rotational symmetry.

At an intermediate distance, after the peak, the force crosses zero.
We identify this distance with $\rt$ and we conservatively define an uncertainty related to the interval containing the first point where the force changes sign from positive to negative.
As shown in Fig.~\ref{fig:rcrh}, $\rt$ defined this way behaves linearly with the temperature $\rt \sim T$ at high temperature.
By definition, $\rt$ cannot be smaller than $\rp$.
In Fig.~\ref{fig:rcrh}, $\rt$ goes closer to $\rp$ as the temperature decreases and it is consistent with $\rt=\rp\simeq \rc$ at $T=0$.

At $r_0>\rt$, the force is repulsive and we will comment on the implications of this below. 
After the repulsive region, at very large $r_0$, we expect $F/(N-1) \sim 1/r_0^8$. 
However, our data is not precise enough to distinguish this from zero. 

\begin{figure}[htbp]
\begin{center}
\rotatebox{0}{
\scalebox{0.3}{
\includegraphics{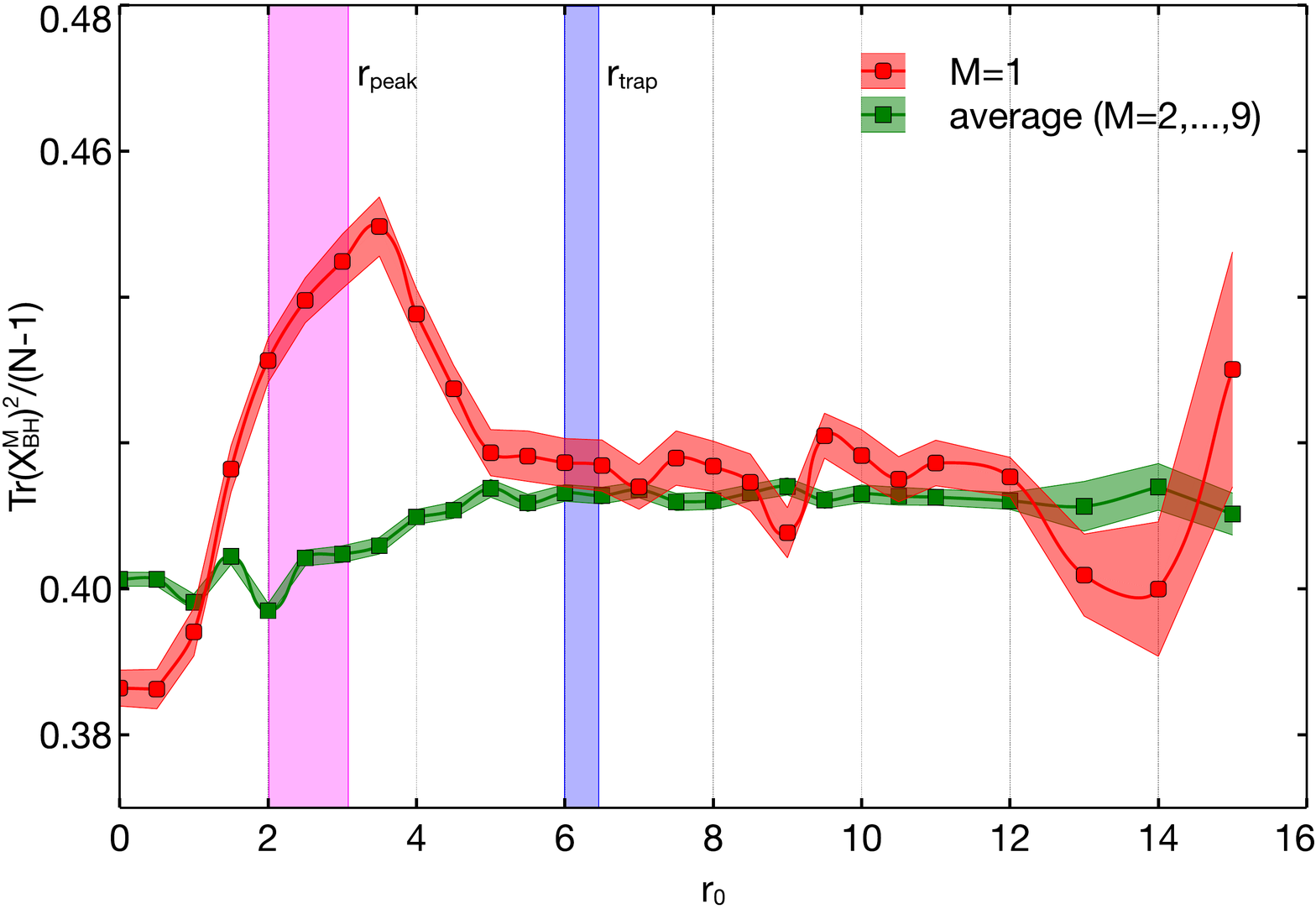}}}
\end{center}
\caption{
The squared radius of the bunch in the direction of the probe $r^{2}_{\textrm{M=1}}=\langle\frac{1}{(N-1)}{\rm Tr}(X_{\rm BH}^1)^2\rangle$ (red) and the squared radius averaged over the eight orthogonal directions, $r^{2}_{\textrm{average}}=\langle\frac{1}{8(N-1)}\sum_{M=2}^9{\rm Tr}(X_{\rm BH}^M)^2\rangle$ (green) as a function of the probe position $r_0$.
The radius of the bunch in M=1 is larger than the one in the orthogonal directions once the probe enters $\rt$ and grows to a maximum near $\rp$.
$\rt$ and $\rp$ are indicated by vertical colored bands, while $r^{2}_{\textrm{M=1}}$ and $r^{2}_{\textrm{average}}$ are shown with error bands representing statistical uncertainties of the Monte Carlo simulations.
The data is for $N=12$, $L=10$ and $T=2.0$, but similar features are present for all parameters $N$,$L$ and $T$ that we studied.
The larger error bands on the $M=1$ direction compared to the orthogonal direction reflects the fact that there are 8 orthogonal directions so we effectively get a larger statistical sample for the orthogonal directions.
}\label{fig:rbunch-shape}
\end{figure}

In Fig.~\ref{fig:rbunch-shape} we plot the square radius of the bunch in the direction of the probe $r^{2}_{\textrm{M=1}}=\langle\frac{1}{(N-1)}{\rm Tr}(X_{\rm BH}^1)^2\rangle$ and the one averaged over the orthogonal directions ($M=2\ldots9$).
We note that, when the probe is far away, the two radii are consistent, while $r^{2}_{\textrm{M=1}}$ quickly grows to a maximum when the probe moves between $\rt$ and $\rp$.
This can be interpreted as a deformation of the bunch due to the interactions with the probe similar to a tidal effect; in fact, when the force has a peak at $\rp$, the bunch is quite prolate.
When $r_0 \lesssim \rp$, the bunch relaxes back to a spherical shape and ultimately becomes oblate as $r_0$ vanishes, although one should take care in this regime, as the geometrical interpretation becomes obscure.

\begin{figure}[htbp]
\begin{center}
\rotatebox{0}{
\scalebox{0.25}{
\includegraphics{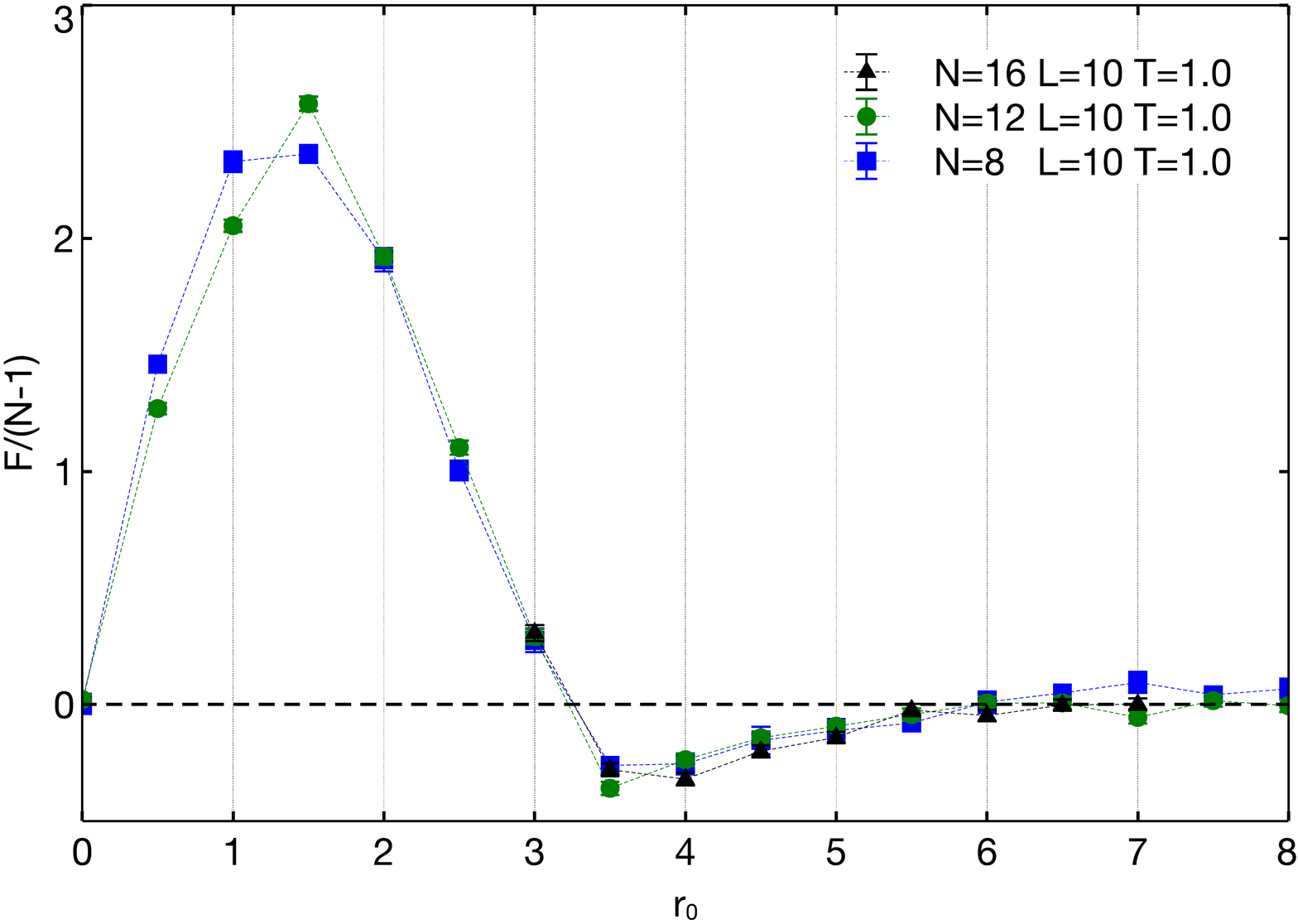}}}
\rotatebox{0}{
\scalebox{0.25}{
\includegraphics{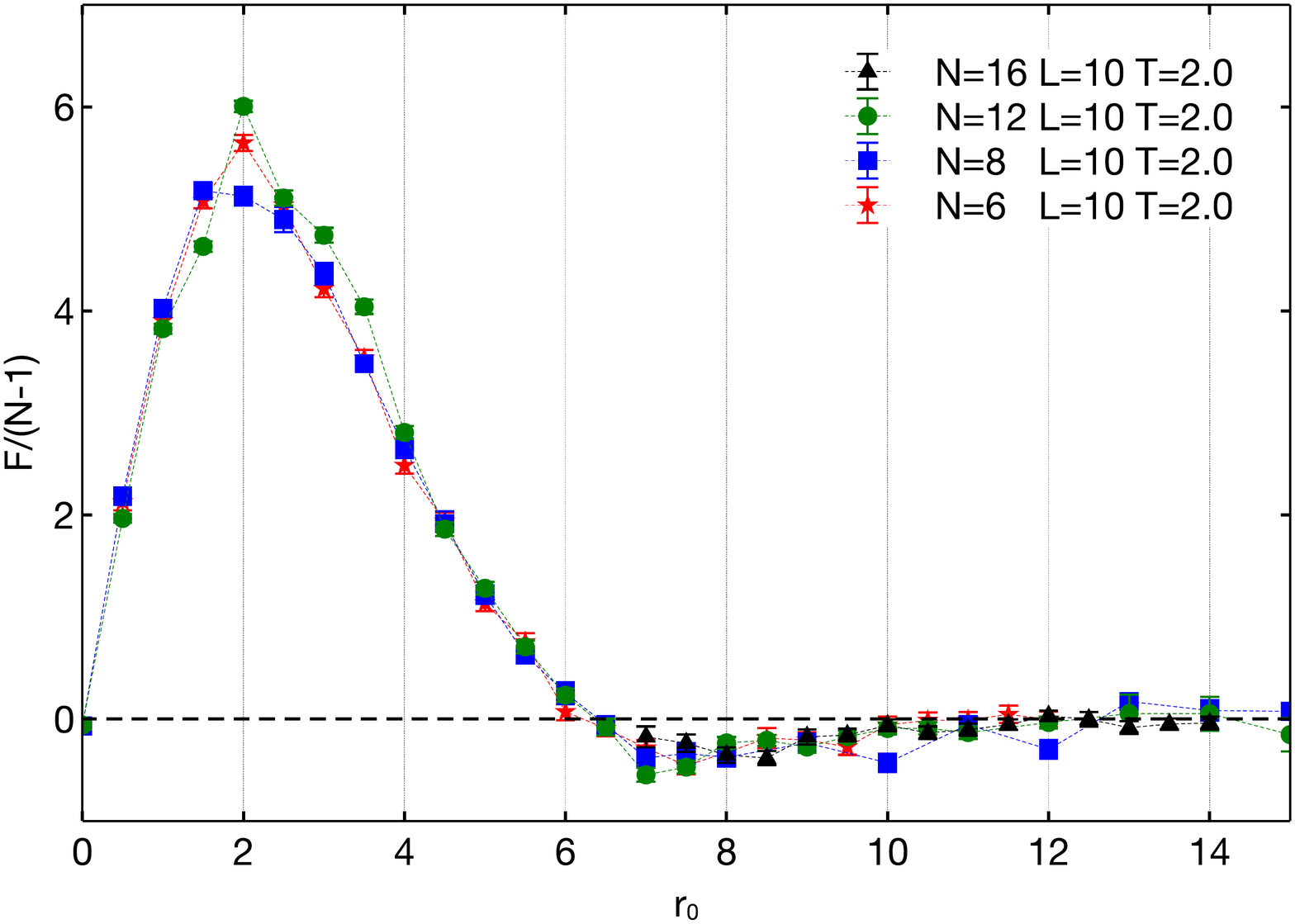}}}
\end{center}
\caption{
$F(N,r_{0})/(N-1)$ at $T=1.0$ and $T=2.0$. For the largest value of $N$ we only have measurements at large $r_{0}$, beyond the region of the peak.}
\label{fig:force_plot}
\end{figure}

\begin{figure}[htbp]
\begin{center}
\rotatebox{0}{
\scalebox{0.3}{
\includegraphics{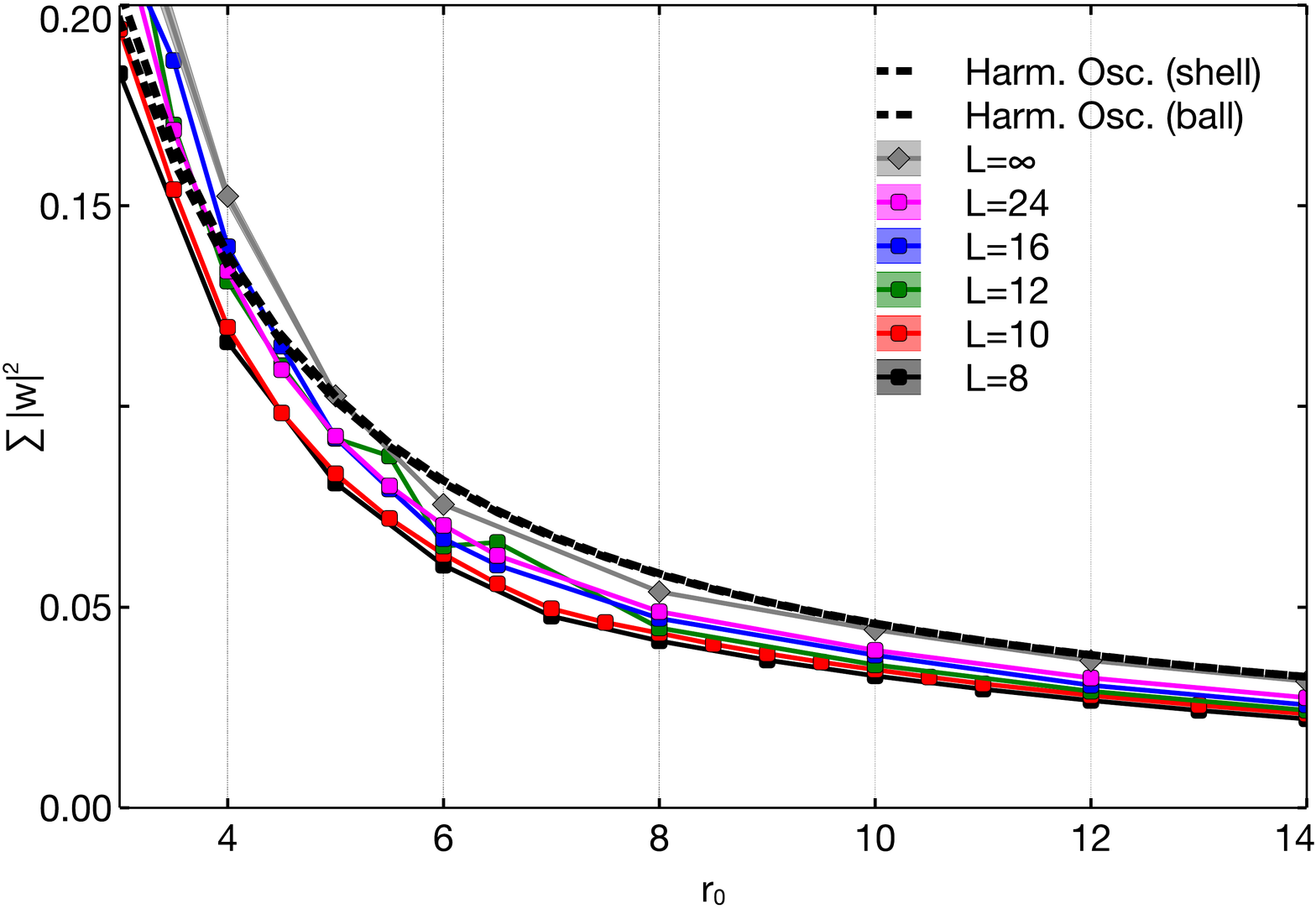}}}
\end{center}
\caption{
Values of $\left(\frac{1}{8\beta}\sum_{M=2}^9\int dt |w_M|^2\right)$ measured on the lattice as a function of $r_{0}$ for $N=12$, $T=2$ and various $L$.
The continuum limit is obtained by extrapolating the finite-$L$ points to $L=\infty$ at each $r_0$ where enough values for a robust estimate are present.
Even when we can not take a reliable continuum limit, we show the fixed-$L$ results.
The perturbative curve is obtained following the procedure in Appendix~\ref{sec:harmonic-oscillator} with $\rc=1.96$, $N=12$ and $T=2$.
The continuum curve at $r_0>10$ is agreeing nicely with the perturbative expectation, while an enhancement can be seen at smaller $r_0 \lesssim 5$.
}\label{fig:w2-T20-N12}
\end{figure}

\begin{figure}[htbp]
\begin{center}
\rotatebox{0}{
\scalebox{0.3}{
\includegraphics{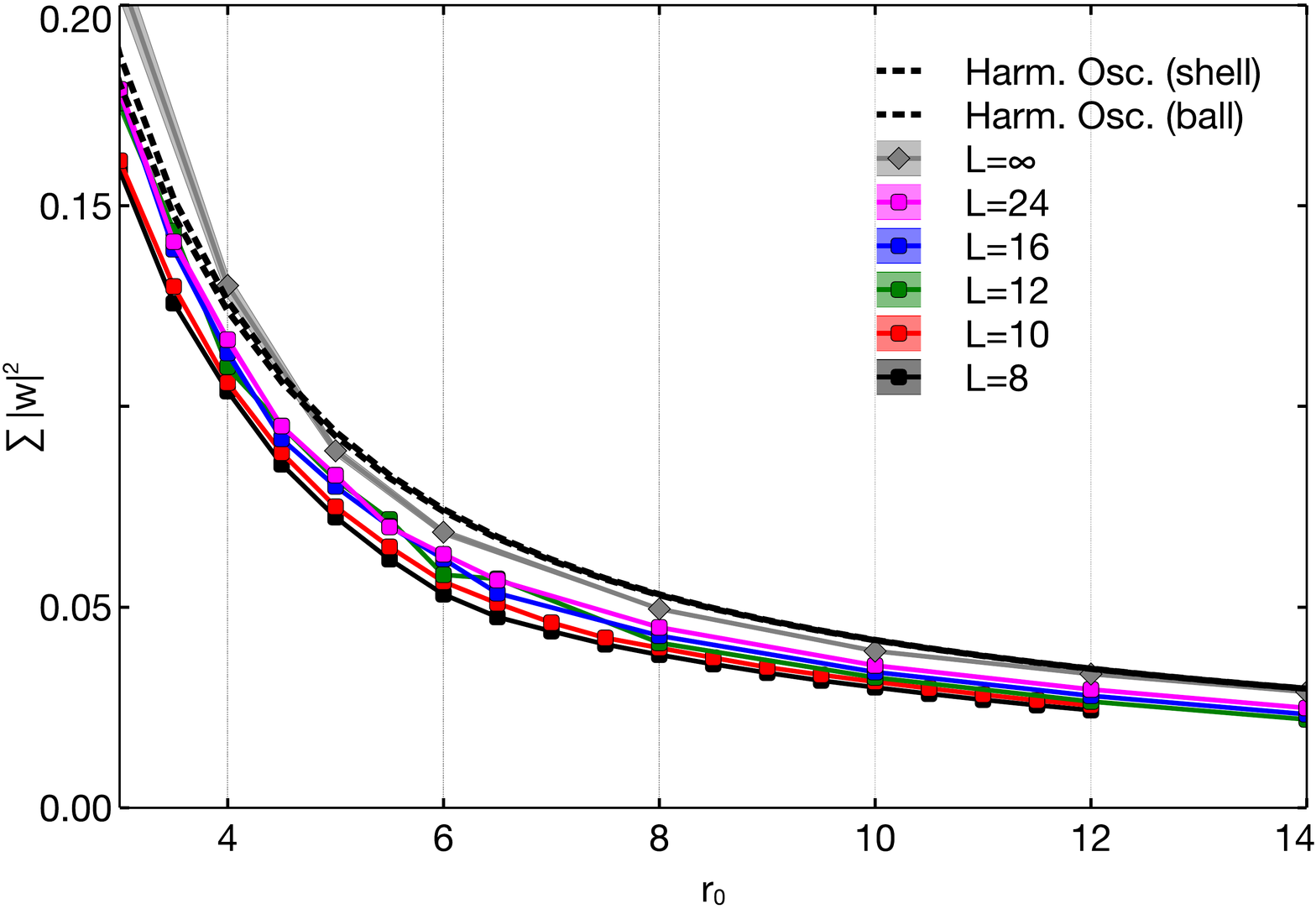}}}
\end{center}
\caption{
Vaules of $\left(\frac{1}{8\beta}\sum_{M=2}^9\int dt |w_M|^2\right)$ measured on the lattice as a function of $r_{0}$ for $N=6$, $T=2$ and various $L$.
The continuum limit is obtained by extrapolating the finite-$L$ points to $L=\infty$ at each $r_0$ where enough values for a robust estimate are present.
Even when we can not take a reliable continuum limit, we show the fixed-$L$ results.
The perturbative curve is obtained following the procedure in Appendix~\ref{sec:harmonic-oscillator} with $\rc=1.91$, $N=6$ and $T=2$.
The continuum curve at $r_0>12$ is agreeing nicely with the perturbative expectation.
An enhancement can be seen at $r_0 \lesssim 5$, though it is less clear compared with $N=12$.
}\label{fig:w2-T20-N6}
\end{figure}

Next, let us consider the size of the fluctuation of the off-diagonal elements, 
\begin{equation}
    \sum|w|^2 \equiv \frac{1}{8\beta}\sum_{M=2}^9\int dt |w_M|^2.
\end{equation}
When $|w_M|$ is small enough that ${\cal O}(|w|^3)$ and ${\cal O}(|w|^4)$ terms in the Lagrangian are negligible, the off-diagonal elements behave as harmonic oscillators.
In this case, $\sum|w|^2$ becomes $\frac{N-1}{2rN}\,\frac{1+e^{-r/T}}{1-e^{-r/T}}$ (for the derivation see Appendix~\ref{sec:harmonic-oscillator}).
Note that we have treated the length of all open strings, connecting the probe brane and the bunch of eigenvalues, to be $r$; this is valid only when $r$ is sufficiently larger than $\rc$.
When taking into account the the finite extent of $\rc$, the harmonic oscillator formula should be replaced by $\sum_{i=1}^{N-1}\frac{1}{2r_{i}N}\frac{1+e^{-r_i/T}}{1-e^{-r_i/T}}$, where $r_i$ is the distance between $i$-th D0-brane in the bunch and the probe. 
Of course, due to the non-commutativity of the matrices, the ``positions'' of D0-branes, and hence the distances, are ambiguous; see \cite{Azeyanagi:2009zf} for detailed argument with numerical inputs. 
Here, for simplicity, we consider a 9-dimensional spherical surface (shell) and a 9-dimensional spherical volume (ball) of radius $\rc$.\footnote{Adding the probe brane breaks the  SO(9) symmetry.  At each $N,L$ and $T$, we could use samples with the largest values of $r_0$,  where the SO(9) symmetry is almost restored, to determine $\rc$. In the following, for our plots at $T=2.0$, we used the extrapolated continuum limit value $\rc=1.96(6)$ for $N=12$ and $\rc=1.91(1)$ for $N=6$.}

In Fig.~\ref{fig:w2-T20-N12}, the values of  $\sum|w|^2 $ as a function of $r_{0}$ are plotted together with the harmonic oscillator value estimated by including the effects of the bunch and of thermal fluctuations.
A continuum limit is also performed by using simulations at different lattice spacings, from $L=8$ to $L=24$.
For some values of $r_0$ we are unable to reliably determine the continuum limit, but we still plot the individual results at fixed lattice spacings.
First of all, we can see that $\sum|w|^2$ in the continuum limit is perfectly consistent with the harmonic oscillators behavior at $r_0 \geq 10$.
At smaller $r_0$ distances the off-diagonal fluctuations become larger than the perturbative estimate, which means many open strings are excited and non-perturbative effects are becoming important.

We emphasize that the notion of ``the position of the probe'' becomes obscure when open strings are non-perturbatively excited.
Fig.~\ref{fig:w2-T20-N12} and Fig.~\ref{fig:w2-T20-N6} suggests that the ``geometry'' becomes obscure approximately at $r<\rt$.
At $r<\rc$, the ``position'' does not even make sense approximately.

\subsection{Comments on the D0/D4 System}\label{sec:D0-D4}
\hspace{0.25in}

The setup discussed above resembles the Berkooz-Douglas matrix model \cite{Berkooz:1996is}, which consists of the D0-brane matrix model plus a flavor sector which describes the open strings stretched between D0-branes and D4-branes.
This flavor sector is analogous to the off-diagonal elements in our D0-brane probe setup. 
The mass of the strings is the distance between D0-branes and D4-branes, which is analogous to the distance between the bunch and the probe in our setup.   
The dual gravity picture is similar to the D3/D7 system \cite{Karch:2002sh} which is often used to study flavor dynamics in AdS/CFT.

This D0/D4 case has been studied in a series of papers \cite{Filev:2015cmz}.
The gravity analysis suggests that, like in the D3/D7 case, a phase transition takes place when the D4 comes close to the BH and touches the horizon; see e.g. \cite{Babington:2003vm,Hoyos:2006gb,Mateos:2006nu}. 
The large-mass (long-distance) and small-mass (short-distance) regions are ``deconfined'' and ``confined'' phases, respectively. 
(In the holographic QCD setup by the D3/D7, the gluons are always deconfined, but quarks can still have a confined phase.)
The order parameter is the condensation of the strings and, in the confined phase, strings are highly excited.    

In \cite{Filev:2015cmz}, some gauge theory results based on Monte Carlo simulations are also shown. 
They did not find a nice agreement with the dual gravity calculation at intermediate distance, but this could be attributed to $\alpha'$ corrections, given their temperature range ($T=1.0$ and $T=0.8$). 

\section{Possible Dual Gravity Interpretations}\label{sec:gravity_dual}
In this section, we discuss what kind of possible dual gravity interpretations can be given for the results or our numerical simulations.
Since we have studied only $T \ge 1$, which is rather high temperature, the dual gravity theory
is expected to suffer from large stringy corrections.
Hence the geometric interpretations simply inspired by supergravity may not be appropriate.
In spite of this possible shortcoming, let us review the standard duality picture and discuss alternative interpretations of the emerging geometry.

\subsection{The Standard Duality Dictionary}
\hspace{0.25in}
 
In this paper, we consider the finite temperature dynamics near the 't Hooft limit ($N\to\infty$ with $\lambda=g_{YM}^2N$ fixed), to which the interpretation in the context of the gauge/gravity duality \cite{Itzhaki:1998dd} can be applied. 
When all $N$ eigenvalues are clumped up to form a bunch, the dual geometry is the near-extremal, near-horizon limit of the type IIA black zero-brane, whose metric in string frame is given by 
\begin{eqnarray}
ds^2
=
\alpha'\left\{
-
\frac{U^{7/2}\left(1-\frac{U_0^7}{U^7}\right)}{\sqrt{240\pi^5\lambda}}dt^2
+
\frac{\sqrt{240\pi^5\lambda}}{U^{7/2}\left(1-\frac{U_0^7}{U^7}\right)}dU^2
+
\sqrt{240\pi^5\lambda}U^{-3/2}d\Omega_8^2
\right\},  
\label{D0-metric}
\end{eqnarray}
where $U$ is the radial coordinate times $(\alpha')^{-1}$, which has the dimension of $[{\rm mass}]$, and $U_0$ is the horizon.
The 't Hooft coupling $\lambda$ has the dimension of $[{\rm mass}]^3$. 
Note also that the curvature radius of $\mathbb{S}^8$ depends on the radial coordinate. 
The dilaton depends on the radial coordinate as well,
\begin{eqnarray}
e^{\phi}
=
\frac{4\pi^2\lambda}{N}
\left(
\frac{240\pi^5\lambda}{U^7}
\right)^{3/4}. \label{D0-dilaton}
\end{eqnarray}
The Hawking temperature is given by 
\begin{eqnarray}
T
=
\frac{7U_0^{5/2}}{16\pi^3\sqrt{15\pi\lambda}} 
\label{Hawking-T}
\end{eqnarray}
and is identified with the temperature of the matrix model. 

The energy of the black hole at finite temperature has been studied numerically on the matrix model side starting in \cite{Anagnostopoulos:2007fw}; 
see also \cite{Catterall:2008yz,Hanada:2008ez,Hanada:2013rga,Kadoh:2015mka,Hanada:2016zxj,Berkowitz:2016jlq}. 
Recent Monte Carlo results in the continuum and infinite-$N$ limit  \cite{Berkowitz:2016jlq} strongly support the validity of the duality, including the string corrections.

From \eqref{D0-metric} and \eqref{Hawking-T}, we can see that the horizon shrinks in the string frame when the effective dimensionless temperature $\lambda^{-1/3}T$ is large, while the horizon expands in Einstein frame due to the non-trivial behavior of the dilaton \eqref{D0-dilaton}.  
Therefore, the $\alpha'$-corrections become larger at higher temperature. 
At $T=0$, the black zero-brane is extremal, i.e. the horizon and the singularity coincide.
However, note that in the 't Hooft limit, $N\to\infty$ is taken before $T\to 0$, and then in Einstein frame there is a parametrically large separation between the singularity and horizon. 
Note that the horizon scales as $U_0\sim T^{2/5}$. 

The probe brane is believed to be described by the Dirac-Born-Infeld (DBI) action in this spacetime.

\subsection{Where is the Horizon?}
\hspace{0.25in}
Now we discuss a few possible geometric interpretations and their advantages, disadvantages, and falsifiability. 
Before going into the details, let us clarify the assumptions regarding the holographic dictionary. 
Firstly, the duality between the real-time theories is only employed without the deformation term, $\Delta \mathcal{L}$. 
The deformation $\Delta \mathcal{L}$ was employed on the gauge theory side just as a numerical trick to determine the force of the original theory in Minkowski signature without the deformation.
In order to relate the original theory with Minkowski signature to string theory, we need neither the deformation nor the Euclidean theory.  

On the gravity side, we consider the motion of a probe D0-brane in the black zero-brane geometry. 
As can be seen from the arguments and calculations in the previous sections, we have assumed that the $(N-1)\times(N-1)$ block $X_{\rm BH}$ corresponds to the black hole, and the probe D0-brane corresponds to the ($N,N$)-components of the matrices.
We assumed ${\rm Tr} X_{\rm BH}/(N-1)$ is the `center' of the black hole, in the sense that the distance between the probe and the black hole is defined by $|{\rm Tr} X_{\rm BH}/(N-1)-x_{\rm D0}|$.
This interpretation can be made precise as long as the stringy effects are not too large;
when many of the open strings are excited (correspondingly, when the $N$-th row and column take large values), the notion of the localized probe becomes obscure. 

We have calculated the force when the relative velocity between the black hole and the probe is zero. 
Without knowing the velocity dependence, we cannot follow the motion of the probe precisely.  
Below we will assume that the velocity dependence does not change the behavior of the system drastically.

For sake of clarity, let us repeat here an important difference between the two temperature regions, $T\gtrsim 1$ and $T\lesssim 1$, which we have briefly mentioned in Sec.~\ref{sec:proposal}.
On the gauge theory side, there are two different sources of the non-commutativity: the thermal excitations and the zero-point oscillations.
The former corresponds to the actual stringy excitations on the gravity side, while the latter may not invalidate the classical gravity picture based on the smooth geometry. 
These two contributions should become of the same order at $T\sim 1$.
Our simulations have been performed for $T \gtrsim 1$, where the bunch is dominated by the thermal excitations.
Below, we will consider $T \gtrsim 1$ in detail, and then briefly comment on $T\lesssim 1$. 

\subsubsection{$T\gtrsim 1$: Is $0\le r\le \rc$ the Horizon?}\label{sec:r<rc}
\hspace{0.25in}
Probably the most conservative interpretation in this high-temperature regime is that the entire bunch, $0\le r\le \rc$, describes the horizon of the type IIA black zero-brane.  
If one believes that all the information about the black hole is encoded in the horizon, why don't we regard the entire bunch, which is the carrier of the information on the gauge theory side, with the horizon? 
This interpretation has some other advantages as well:
\begin{itemize}
\item
If the gauge theory describes the system from the exterior observer's viewpoint, the light modes should appear near the horizon due to the redshift. 
The light strings between the bunch and the probe, which become massless when the probe reaches $r=\rc$, are natural counterparts.
See \cite{Ferrari:2016bvq} for a related consideration for a solvable model. 

\item
On the gravity side, the dynamics at the horizon naturally explains fast scrambling \cite{Sekino:2008he}. 
On the gauge theory side, the non-local interaction mediated by open strings is crucial for fast scrambling. 
Then $0\le r\le \rc$, where the open strings condense, is a natural place where fast scrambling can take place.
Note however that this argument may not exclude the possibility that $r=\rt$ is the horizon, because the open string excitations are enhanced for $r\le \rt$.

\item
In this interpretation, the interior of the horizon cannot be seen from the eigenvalue distribution. 
This is an advantage when we consider the theory with Euclidean signature, whose gravity dual does not have an `interior'. 

\end{itemize}

A possible difficulty of this interpretation is that the physical meaning of the distance scale $\rt$ is not clear. 
It may not be an immediate problem, especially in the D0-brane case, in which the $\alpha'$-corrections are inevitable at finite temperature. 
We will come back to this point later. 
Note also that this difficulty might be seen as good news because, in case one determines the existence/absence of such distance scale by studying the dynamics of the probe D-brane from string theory, it is possible to test this interpretation. 

In Ref.~\cite{Berenstein:2013tya}, the spectrum of the Dirac operator acting on the fermion $\psi$ has been studied by means of semi-classical simulations. 
In such simulations, a $(N-1)\times (N-1)$ matrix $X_{\rm BH}^M$ was generated and a `probe D0-brane' $x_{\rm D0}^M$ was introduced by hand.
The spectrum of the Dirac operator obtained from the matrix
\begin{eqnarray}
X^M
=
\left(
\begin{array}{cc}
X_{\rm BH}^M & 0 \\
0 & x_{\rm D0}^M
\end{array}
\right)
\end{eqnarray}
was then studied.
Ref.~\cite{Berenstein:2013tya} identified the horizon with the distance scale where the Dirac operator becomes gapless. 
This length scale is likely to be our $\rc$.
\subsubsection{$T\gtrsim 1$: Is $\rt$ the Horizon?}\label{sec:r=rt}
\hspace{0.25in}
Another possibility is that $\rt$ is the horizon. 
This interpretation has some favorable features:
\begin{itemize}
\item
When $r$ is slightly above $\rt$, the force is repulsive. 
This is not something expected in the interior of the black hole. 
It is natural to regard $r>\rt$ to be (at least a part of) the exterior. 


\item
As mentioned above, the off-diagonal elements are highly excited when $r<\rt$, although they do not condense until the probe goes to $r\le \rc$. 
Such excitations can explain why the D-branes are trapped there; see Sec.~\ref{sec:proposal}.
Furthermore, when a D-brane is emitted to $r>\rt$, the temperature of the black hole goes up \cite{Berkowitz:2016znt,essay}.  

\item
Note also that, if we identify $\rh$ with $\rt$, it is consistent with a conservative stance on possible stringy effects --- if stringy effects should become relevant, it should be at $r\le \rh$. 

\end{itemize}

In \cite{Hyakutake:2013vwa}, the force acting on a D0-brane probe outside the horizon was studied on the gravity side. 
At the level of supergravity, the force is attractive at any distance.
When $O(g_s)$ corrections are taken into account, a repulsive force correction is added near the horizon.
The effects from the $\alpha'$ corrections and the higher order terms in $g_s$ are not known. 
If $\rt$ is the horizon, then our result on the gauge theory side ($O(N)$ repulsion) suggests that the $\alpha'$ corrections lead to a repulsion near the horizon.
It provides us with the falsifiability of this interpretation.\footnote{
We would like to thank Y.~Hyakutake for the discussion concerning this point.} 

The disadvantages of this interpretation include the following:
\begin{itemize}
\item
This distance scale makes sense in the Euclidean theory as well. 
Then this interpretation would mean that the dual Euclidean black hole geometry may somehow knows the black hole interior, which is against usual lore.  

\item
If $\rt$ is the horizon, then the probe can pass through the horizon within a finite time in gauge theory.
If we then identify the time in gauge theory with the exterior observer's time as usual, it suggests that the in-falling observer can go into the black hole within a finite time as seen from the exterior observer's clock.\footnote{
We would like to thank T.~Banks for pointing out this problem. 
}$^,$\footnote{
A possible resolution in the philosophy of the Matrix Theory Conjecture --- everything is made of eigenvalues --- is as follows.
Suppose everything, including the in-falling and exterior observers, is made of eigenvalues. 
They communicate with each other by exchanging eigenvalues. 
As the in-falling observer goes parametrically close to $\rt$, say the distance of order $1/N$, stringy effects turn on and make it hard to send eigenvalues to the exterior observer.
Then the exterior observer would have to wait longer to receive the message. 
}

\end{itemize}
\subsubsection{$T\gtrsim 1$: Is $r \le \rt$ the Horizon?}\label{sec:r<rt}
\hspace{0.25in}
An important and subtle point related to the disadvantages mentioned in the end of Sec.~\ref{sec:r=rt} is that, when $r<\rt$, many strings are excited, hence it is not clear whether a smooth geometry can make sense there. 
(Clearly, for $r\le \rc$, the geometry does not make sense.)
If a smooth geometry does not make sense, the ``interior'' of the black hole may not make sense; 
it would be better to regard the entire region $r<\rt$ to be some `stringy stuff' which represents the horizon.
Then the disadvantages mentioned above can be resolved.

\subsubsection{$T\gtrsim 1$: Fuzzball?}
\hspace{0.25in}
Yet another possibility is the fuzzball (see e.g. \cite{Mathur:2005zp} for a review).
In this interpretation, space itself ends at $r=\rc$ (or $r=\rt$) due to some stringy stuff. 
At this moment we do not know how to distinguish this possibility from the scenarios suggested in Sec.~\ref{sec:r<rc} and Sec.~\ref{sec:r=rt}.

\subsubsection{$T\lesssim 1$: Low-temperature Region}
\hspace{0.25in}
As we have commented before, at low temperature, a large non-commutativity does not necessarily mean the breakdown of smooth spacetime, as long as the thermal excitation on top of the quantum fluctuation is not large; hence the geometry would make sense even at $r<\rc$. 
The results of Ref.~\cite{Iizuka:2001cw} and Ref.~\cite{Azeyanagi:2009zf} seem to be consistent with this expectation. 
In this case it would be natural to expect that the horizon is hidden below $\rc$, as discussed in Ref.~\cite{Iizuka:2001cw}. 
It fits well with the standard duality dictionary, in which the radial coordinate $U$ in 
\eqref{D0-metric} is identified with $r$ up to a constant multiplicative factor. 
Note that the horizon is at $U_0\sim T^{2/5}$, as one can see from \eqref{Hawking-T}.  

As $T$ becomes large, $U_0\sim T^{2/5}$ increases. At $T\sim 1$, 
it can become as large as $\rc$ and $\rt$. 
Hence it would be natural to think that $\rc$ and $\rt$ at high temperature 
is related to the horizon. 
At this moment this is just a speculation as we have yet to study this scenario in the low temperature region.

\section{Discussion}
\hspace{0.25in}
In this paper, we have studied the dynamics of eigenvalues in gauge theories, particularly in the D0-brane matrix model.
We have performed explicit numerical calculations in high temperature region. 
There are two length scales, which we denoted by $\rt$ and $\rc$, which may be related to the horizon on the gravity side. 

Our study of the high temperature region has pros and cons. 
The biggest pro is that the stringy effect is large, and the largest con is that the stringy effect is large. Stringy effects are something we want to learn from gauge theory, but at the same time, 
when the stringy effects are too large the gravity interpretation is not easy.   
As a next step, it is necessary to study a parameter region responsible for small stringy effects.
There are two natural approaches: (1) long distance, corresponding to far outside the bunch, 
regardless of the temperature, and (2) low temperature inside the bunch.   
The former is more straightforward as off-diagonal elements are suppressed. 
(Note that stringy effects are large at long distance, but as long as we only look at the dynamics of the eigenvalues we expect the DBI action to provide an accurate description.)
For the latter, we need to resolve the problem of the non-commutativity. 
One possible approach is to study D0/D4 system (Sec.~\ref{sec:D0-D4}).\footnote{
We would like to thank J.~Maldacena for suggesting this approach.}
Here, the masses of the flavor sector specifies the position of the D4 at spatial infinity, 
and if the critical mass agrees with the dual gravity prediction it means that the D4 probe is actually described by the DBI action. 
We can also use the D1-probe in $(1+1)$-d SYM. Fixing the two end points far outside bunch and allowing the middle of the D1 to fall down into the bunch, the shape of the probe can be determined, 
(at least outside the bunch), and it is possible to test if the DBI action is valid there.

In the correspondence between $(p+1)$-dimensional SYM and the black $p$-brane \cite{Itzhaki:1998dd}, one can probe the geometry in the same way, by using D$p$-branes; see e.g. Ref.~\cite{Ferrari:2012nw}.
An important difference from the case of D0-branes is that various shapes can appear. 
Using lattice simulations, it should be possible to see a minimal surface directly. 
Other probes such as the D-instanton can also be useful.  
An important point, which is not apparent from the current analysis, is whether or not the horizon depends on the kind of probe.

In the D0-brane quantum mechanics, the Schwarzschild black hole in eleven dimensions is expected to emerge in the M-theory parameter region, which is at much lower temperatures than the 't Hooft large-$N$ limit.
Also, 4D ${\cal N}=4$ SYM on $\mathbb{S}^3$ is expected to contain low-energy states describing the ten-dimensional Schwarzschild black hole \cite{Witten:1998qj}.
On the gauge theory side, they should be described by bunches of eigenvalues of scalar fields (see e.g. Refs.\cite{Banks:1997hz,Hanada:2016pwv}) and hence the method proposed in this paper can be applied; it is very important to study the emergent geometries in these cases. 
Another important direction is the time-dependence; see e.g. Refs.~\cite{Kofman:2004yc,Berkowitz:2016znt,essay} for previous attempts.

\section*{Acknowledgments}
\hspace{0.25in}
The authors would like to thank O.~Aharony, T.~Banks, D.~Kabat, F.~Ferrari, G.~Lifschytz, J.~Maldacena, E.~Martinec, Y.~Nomura, D.~O'Connor, K.~Papadodimas, J.~Penedones, S.~Shenker, H.~Shimada, E.~Silverstein, K.~Skenderis, B.~Sundborg, L.~Susskind, and V.~Filev for discussions and comments.
The work of M.~H. is supported in part by the Grant-in-Aid of the Japanese Ministry of Education, Sciences and Technology, Sports and Culture (MEXT) for Scientific Research (No.~25287046 and 17K14285).
The work of J.~M. is supported by the California Alliance fellowship (NSF grant 32540).
This work is supported in part by the DFG and the NSFC through funds provided to the Sino-German CRC 110 ``Symmetries and the Emergence of Structure in QCD'' (EB).
E.~R. is supported by a RIKEN Special Postdoctoral fellowship.
This work was performed under the auspices of the U.S. Department of Energy by Lawrence Livermore National Laboratory under contract~{DE-AC52-07NA27344}.
Computing was provided through the Lawrence Livermore National Laboratory (LLNL) Institutional Computing Grand Challenge program.

\appendix
\section{Numerical Calculation of the Force}\label{sec:force-calculation}
\hspace{0.25in} In this appendix we present a typical example of the numerical Monte Carlo simulations described in Sec.~\ref{sec:numerics}.
We take a representative set of parameters $\{N,L,T\} = \{8,8,3.0\}$ where we have explicitly checked how the force $F(N,r_0;c) = 2c(r_0-r)$ depends on $c$, according to the potential in \eqref{eq:deformation}.

\begin{figure}[htbp]
\begin{center}
\rotatebox{0}{
\scalebox{0.3}{
\includegraphics{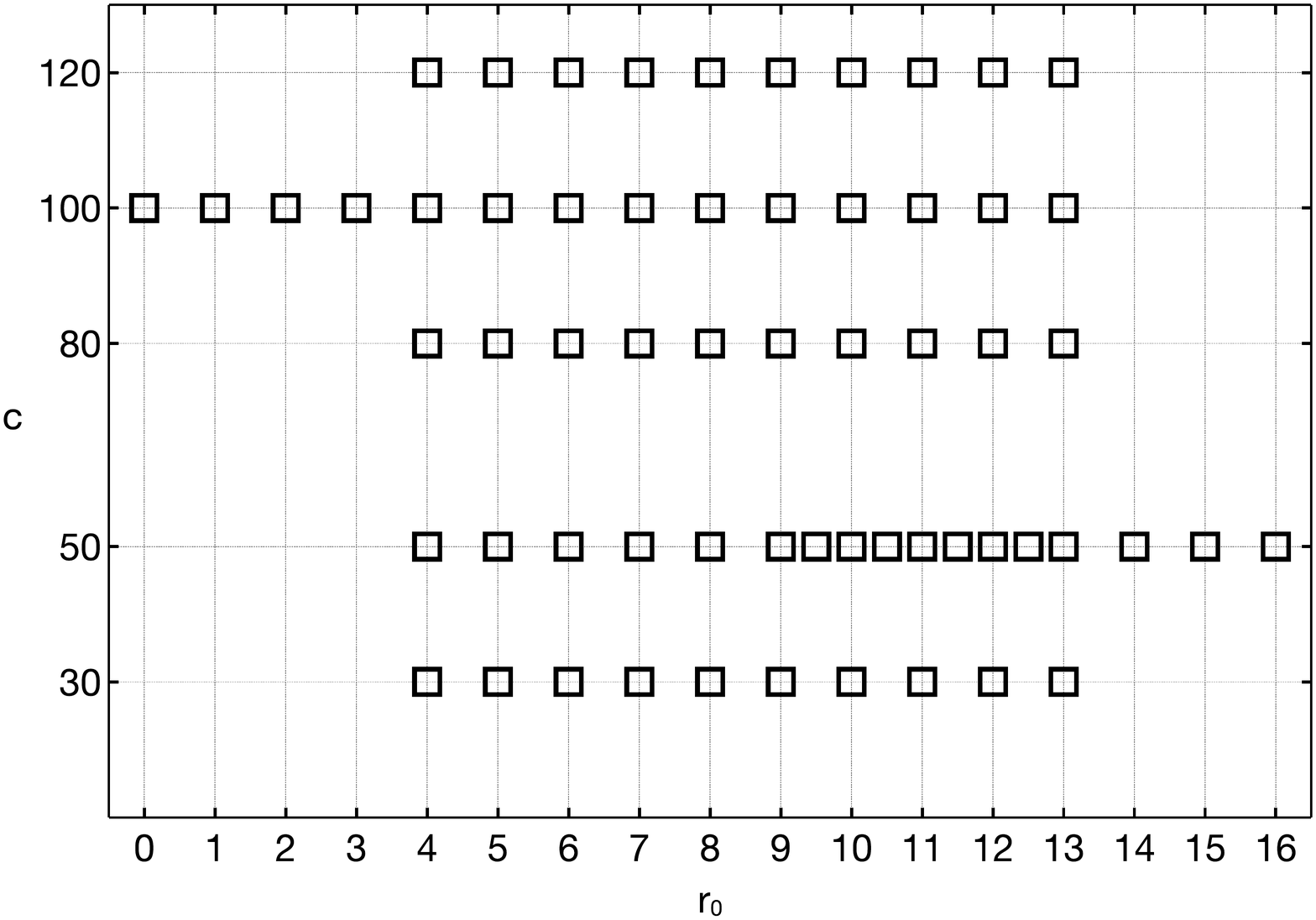}}}
\end{center}
\caption{Map of the $c$ values used at each $r_0$ distance for $N=8$, $L=8$ and $T=3.0$. The distance where the force becomes negative is $r_0\in[9,10]$.}
\label{fig:c-values}
\end{figure}

For $N=8$, $L=8$ and $T=3.0$ we have used several values of $c$, going from 30 to 120, across the whole range of ``constraining'' distance $r_0$.
Remember from \eqref{eq:deformation} that $r_0$ is the distance along direction $M=1$ where the constraining potential is centered, while $c$ is the strength of the quadratic potential.
A summary of the values of $c$ used for this point in parameter space as a function of $r_0$ is shown in Fig.~\ref{fig:c-values}.

\begin{figure}[htbp]
\begin{center}
\rotatebox{0}{
\scalebox{0.25}{
\includegraphics{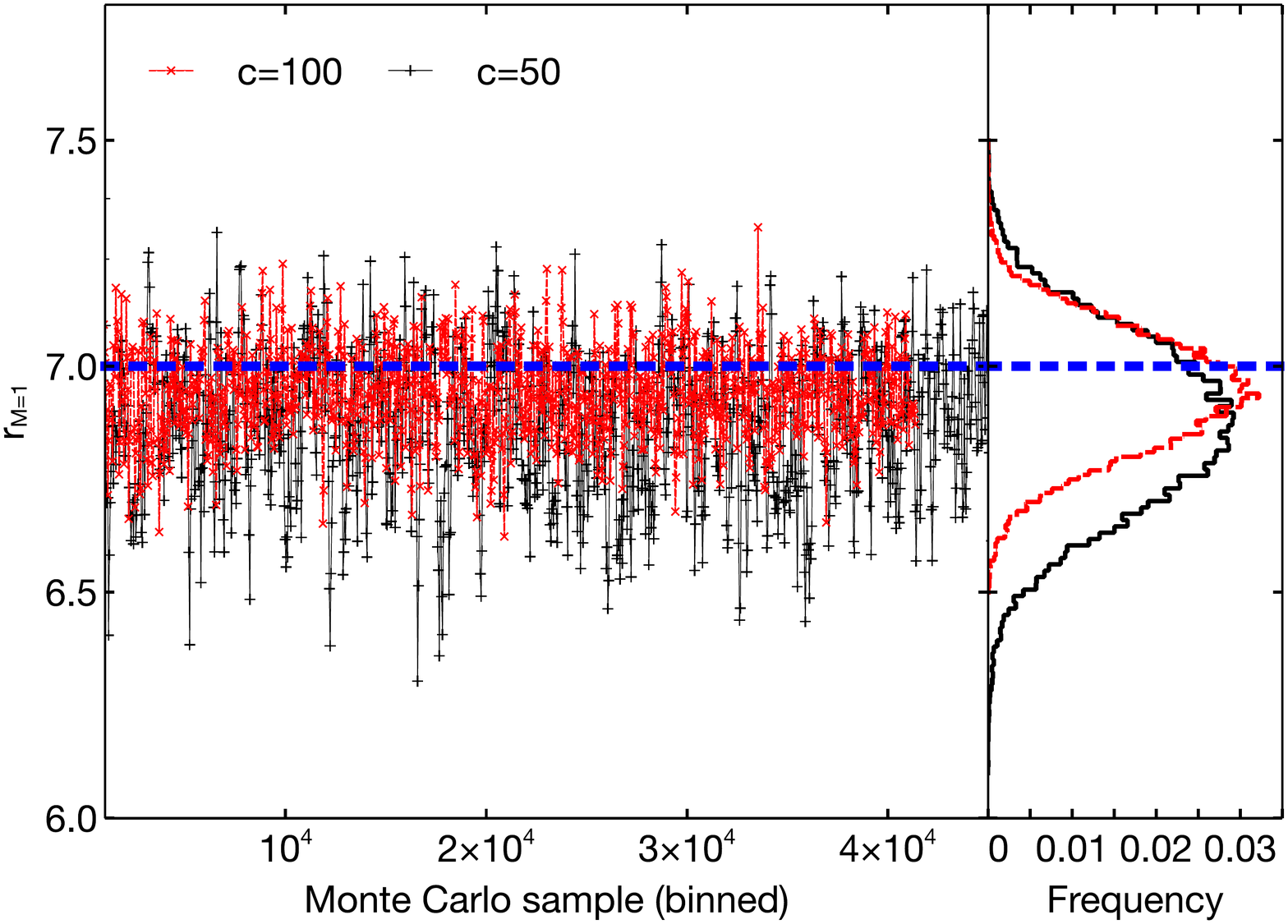}}}
\rotatebox{0}{
\scalebox{0.25}{
\includegraphics{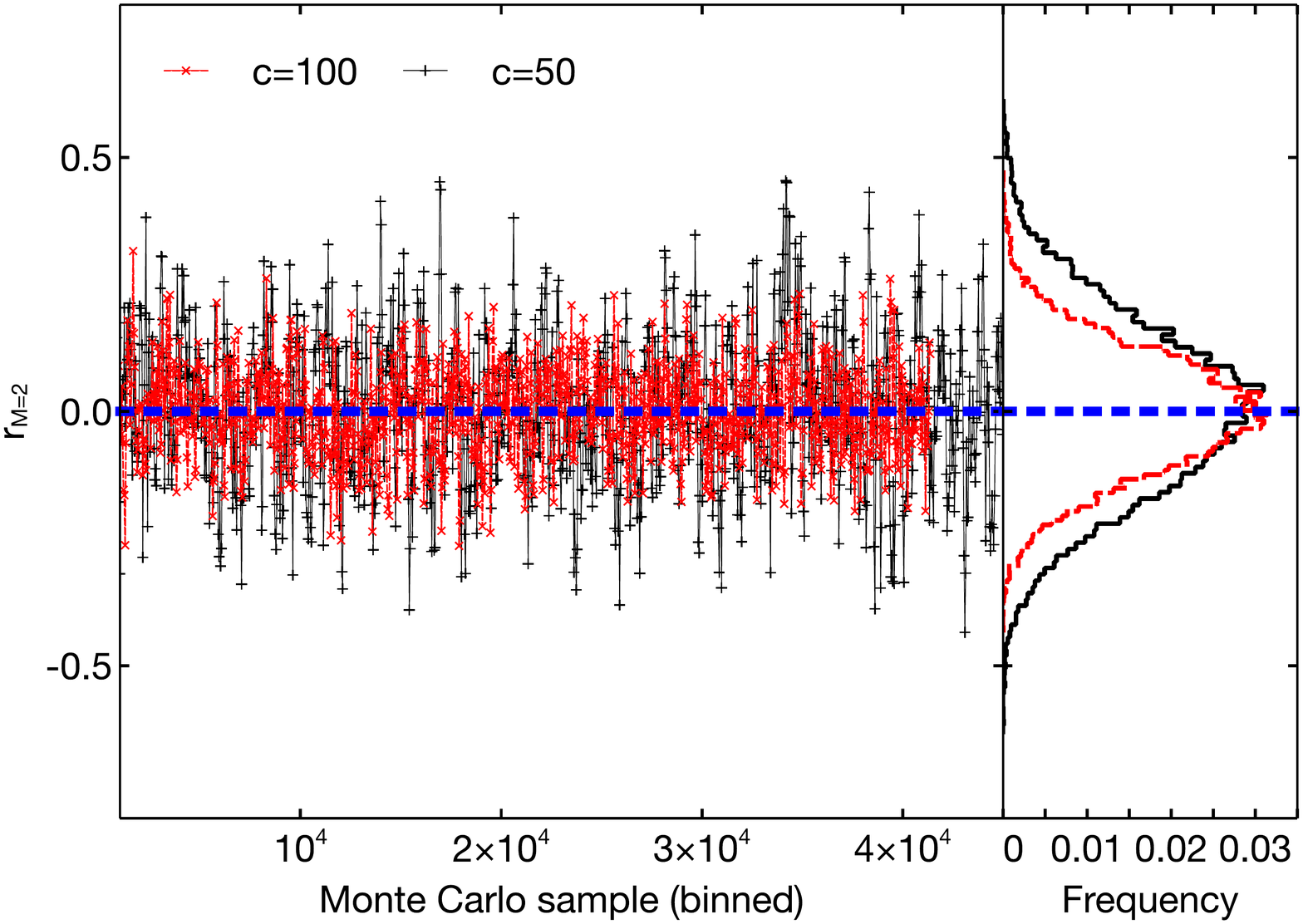}}}
\end{center}
\caption{
Monte Carlo history (binned in blocks of 40 samples for clarity) and histogram of the two observables $r_{M=1}$ and $r_{M=2}$ defined in \eqref{eq:observables} at $r_0=7.0$, for $N=8$, $L=8$ and $T=3.0$. On each plot we show two values of $c$, $c=50$ and $c=100$. The blue dotted line in the left plot corresponds to $r_0=7$, while on the right plot it corresponds to zero.}
\label{fig:distance_plot}
\end{figure}

For larger values of $c$, the probe will be more tightly constrained around $r_0$ in the $M=1$ direction, and around zero in the perpendicular directions.
We show the observables $r_{M=1}$ and $r_{M=2}$ in the left and right plots of Fig.~\ref{fig:distance_plot}, respectively.
For each panel we report the Monte Carlo history and the histogram of the observables, after the initial 1000 samples are discarded for thermalization.
The plot of $r_{M=1}$ show that, for a potential centered around $r_0=7.0$ in the $M=1$ direction (and for $N=8$, $L=8$ and $T=3.0$) the actual coordinate of the probe in such direction is very close to $r_0$, and more so for larger $c$, as expected.
Similarly for $r_{M=2}$, the distribution of the samples is narrower around zero when $c=100$ rather than $c=50$, again confirming that our constraining potential is behaving correctly.

For both values of $c \in \{50, 100\}$ the force near distance $r_0$ is the same, because the expected shift of the probe $r_0–r$ is smaller for larger $c$, and the two effects cancel in the force $F = 2c(r_0–r)$.
This cancellation will break down if the potential is too shallow or the force is too strong---if the probe wanders far from the center of its potential.
 We observe deviations of this kind, for example, if $c=30$ is used at small $r_0=4.0$ or $r_0=5.0$, where the force is large and positive.
This is clearly exemplified in Fig.~\ref{fig:force_plot}.
Typically, in the region where the force is attractive, between $r_0=0$ and the transition to a repulsive force, we choose $c=100$, while we settle on $c=50$ in the region where the force is small or almost zero.

\begin{figure}[htbp]
\begin{center}
\rotatebox{0}{
\scalebox{0.25}{
\includegraphics{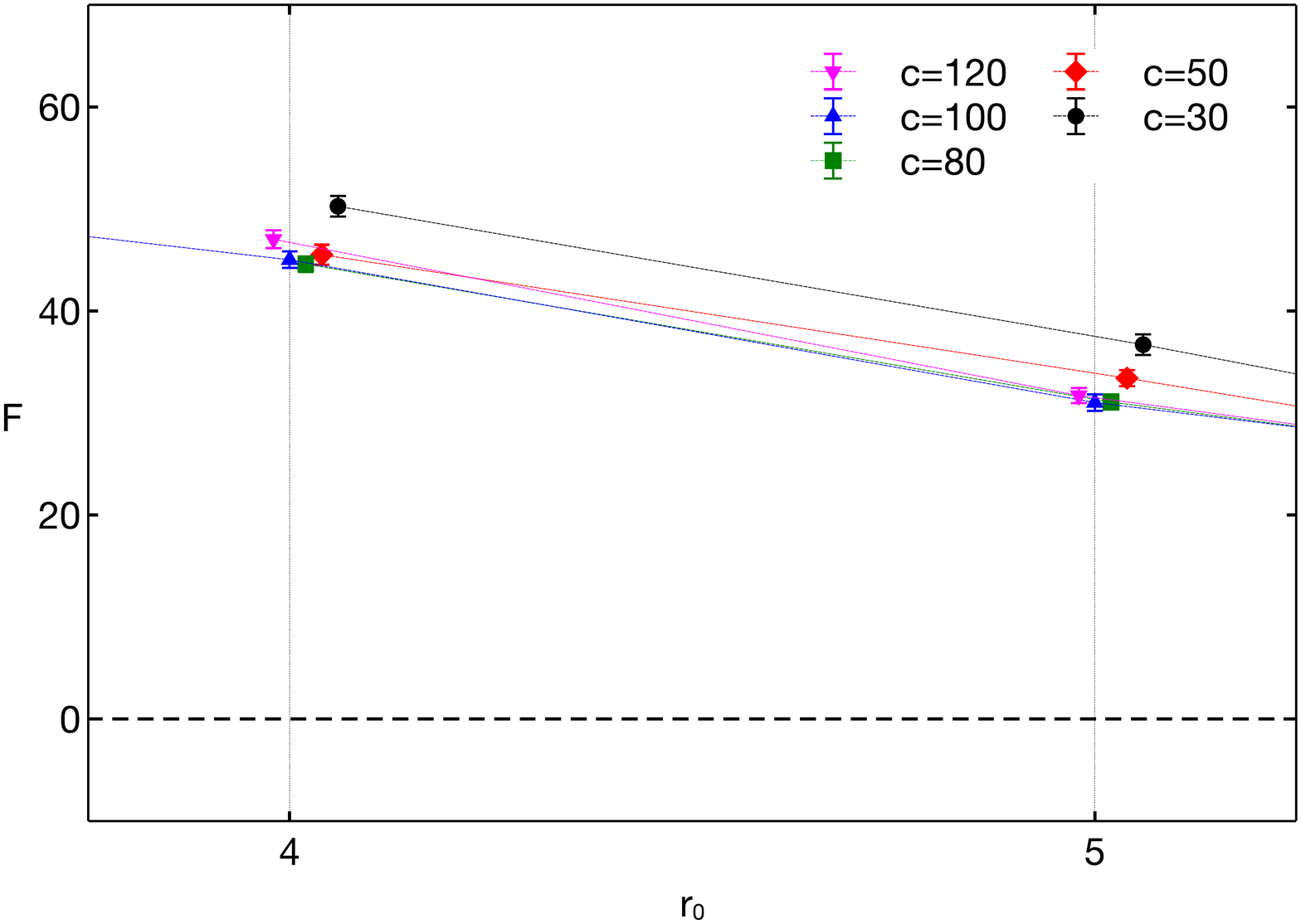}}}
\rotatebox{0}{
\scalebox{0.25}{
\includegraphics{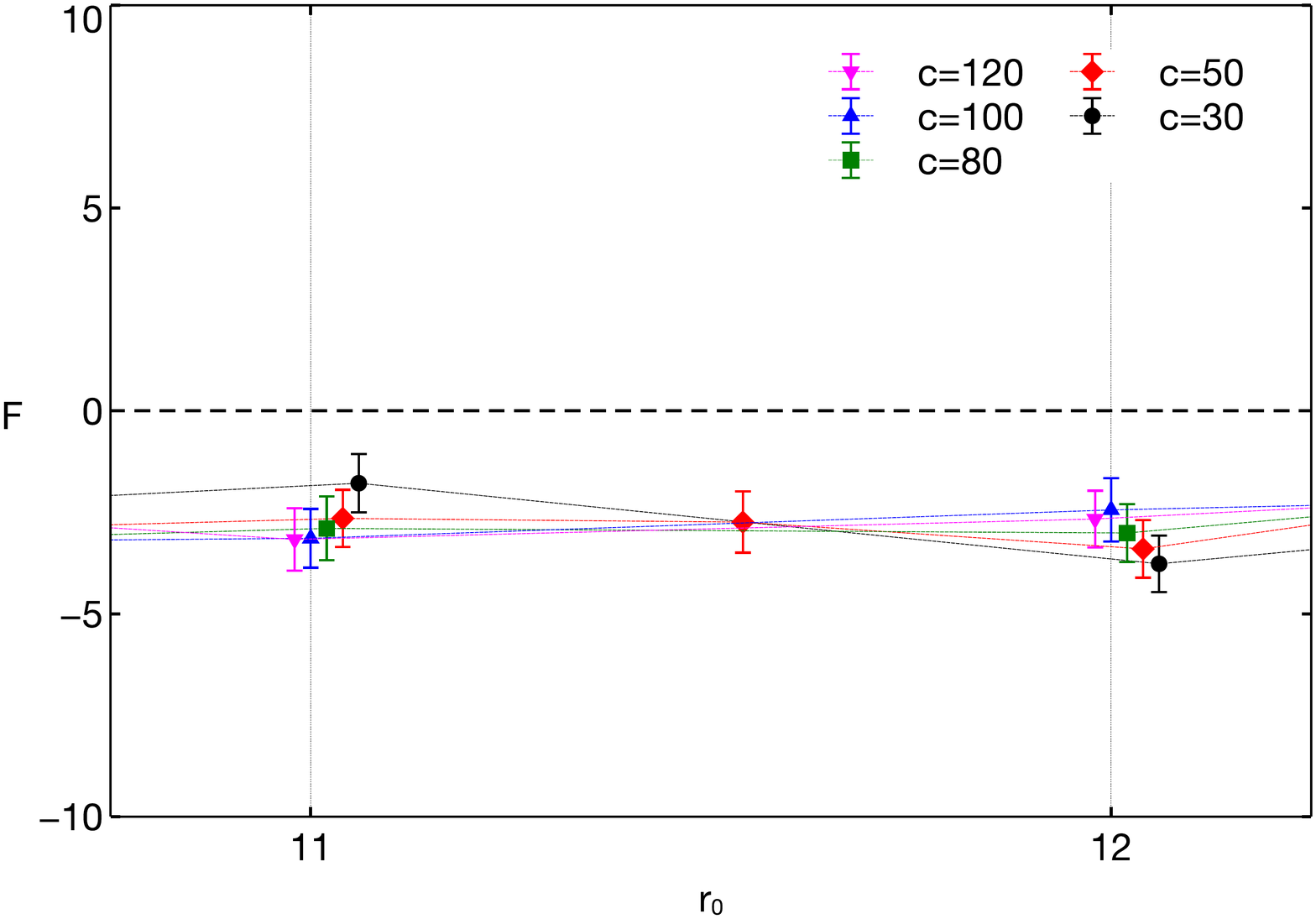}}}
\end{center}
\caption{
Force $F(N,r_0;c)=2c(r_0-r)$ measured for $N=8$, $L=8$ and $T=3.0$ at two distances $r_0$ and five values of $c$. When distance is small (left panel) the value of $c$ for which the force is independent of $c$ must be larger than $c=30$. When the distance is large (right panel) all values of $c$ that we tried are equivalent within the statistical precision of our measurements.}
\label{fig:force_plot}
\end{figure}

\begin{figure}[htbp]
\begin{center}
\rotatebox{0}{
\scalebox{0.3}{
\includegraphics{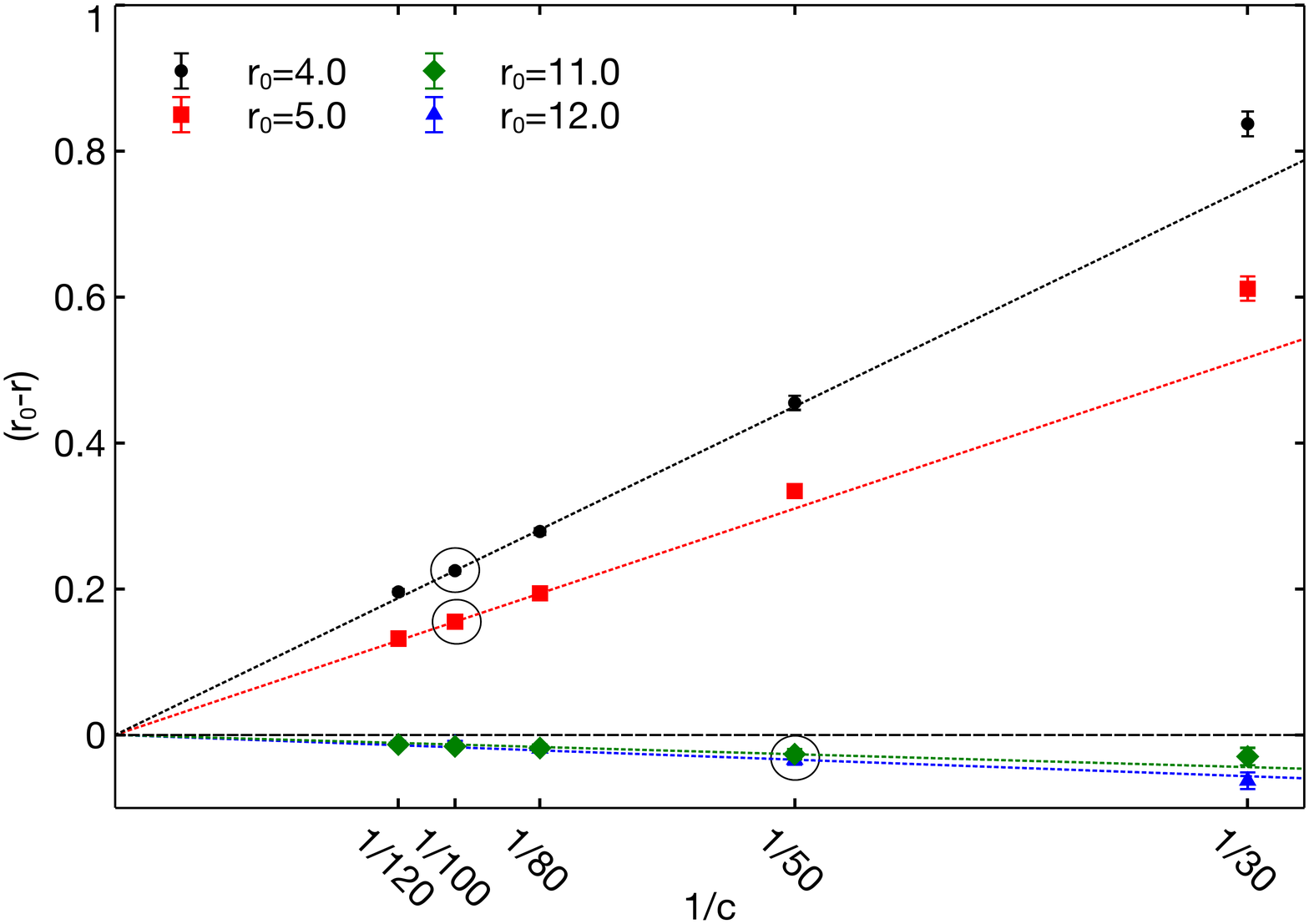}}}
\end{center}
\caption{ For fixed $r_0$ distance at $N=8$, $L=8$ and $T=3.0$, we plot the probe shift $(r_0-r)$ as a function of $1/c$. The slope of the linear relation defines the force $(r_0-r)=F/2 \; 1/c$. A deviation from the linear behavior is present when the force is large and $c$ is too small. The dotted lines represent the slopes for our typical choices of $c=100$ in the small $r_0$ regions and $c=50$ in the large $r_0$ region (circled points).}
\label{fig:force_slope}
\end{figure}

Another equivalent way to look at this is to investigate the relation between $(r_0-r)$ and $c$ at fixed $r_0$: our definition of force will be correct in the region of $c$ where the data is described by a linear function.
We show the probe deviation from the center of the potential $(r_0-r)$ as a function of $1/c$ in Fig.~\ref{fig:force_slope}, for the same four values of $r_0$ reported in Fig.~\ref{fig:force_plot}.
We plot the force obtained from the relation $F=2c(r_0-r)$ with $c$ fixed to the typical values reported above at different $r_0$.
Note again the deviation of the $c=30$ measurements from the linear behavior when the force is large.

\section{Behavior of Off-diagonal Elements}\label{sec:harmonic-oscillator}
\hspace{0.25in}At long distances, the off-diagonal elements are approximated by harmonic oscillators. There are $8(N-1)$ complex d.o.f., 
and hence $16(N-1)$ harmonic oscillators. 
By writing $w_{M,j}=\frac{(x_{M,j}+iy_{M,j})}{\sqrt{2N}}$, the action for the off-diagonal part can be written as 
\begin{eqnarray}
\sum_{M=2}^9
\sum_{i=1}^{N-1}
\left(
\frac{\dot{x}_{M,i}^2}{2}
+
\frac{\dot{y}_{M,i}^2}{2}
+
\frac{r^2x_{M,i}^2}{2}
+
\frac{r^2y_{M,i}^2}{2}
\right)
\nonumber
\end{eqnarray}
up to the higher order terms. Note that we did {\it not} rescale $r$. 
Hence $x$ and $y$ are harmonic oscillators with $m=1$ and $\omega=r$. 
The ground state wave function is $\sim e^{-x^2/2}, e^{-y^2/2}$. 
At sufficiently long distances, $|w|^2$ is approximated by zero-point fluctuations,
and
$\langle x^2\rangle =\langle y^2\rangle \simeq \frac{\int dx x^2 e^{-rx^2}}{\int dx e^{-rx^2}}=\frac{1}{2r}$, 
$\langle |w_{M,i}|^2\rangle \simeq\frac{\langle x^2\rangle+\langle y^2\rangle }{2N}=\frac{1}{2Nr}$. 
Therefore, 
$\frac{1}{8\beta}\sum_{M=2}^9\int dt |w_M|^2\simeq \frac{N-1}{N}\cdot\frac{1}{2r}$.  
When the excited modes are taken into account, this expression is modified to $\frac{N-1}{N}\cdot\frac{1}{2r}\cdot\frac{1+e^{-r/T}}{1-e^{-r/T}}$. 

Because the bunch of D0-branes has finite size, there is a correction to the above expression. 
If one imagines the D0-branes to be distributed in a spherically symmetric manner, the average distance between them and the probe brane is larger than $r$. 
Hence the fluctuation of the off-diagonal elements should be slightly smaller. 
We have numerically implemented two possible distributions for the D0-branes in the bunch to assess the corrections due to non-zero bunch size:
\begin{itemize}
\item an 8-dimensional sphere $\mathbb{S}^8$ of radius $\rc$, 
\item a 9-dimensional ball of radius $\rc$.
\end{itemize}
The bunch effects have been taken into account by replacing $r$ with $r-r_s$, where $r_s$ is the position of a point randomly sampled according to the aforementioned two distributions, and by taking the average of the function over the whole sample.
We have checked that the final result does not change within the needed precision when the number of samples is large enough.


\end{document}